\documentclass[12pt]{article}
\usepackage{latexsym}
\usepackage{amsmath,amsfonts}
\usepackage{times}
\allowdisplaybreaks[4]

\catcode`\@=11

\newcount\hour
\newcount\minute
\newtoks\amorpm \hour=\time\divide\hour by 60\minute
=\time{\multiply\hour by 60 \global\advance\minute by-\hour}
\edef\standardtime{{\ifnum\hour<12 \global\amorpm={am}%
        \else\global\amorpm={pm}\advance\hour by-12 \fi
        \ifnum\hour=0 \hour=12 \fi
        \number\hour:\ifnum\minute<10
        0\fi\number\minute\the\amorpm}}
\edef\militarytime{\number\hour:\ifnum\minute<10
0\fi\number\minute}

\def\draftlabel#1{{\@bsphack\if@filesw {\let\thepage\relax
   \xdef\@gtempa{\write\@auxout{\string
      \newlabel{#1}{{\@currentlabel}{\thepage}}}}}\@gtempa
   \if@nobreak \ifvmode\nobreak\fi\fi\fi\@esphack}
        \gdef\@eqnlabel{#1}}
\def\@eqnlabel{}
\def\@vacuum{}
\def\marginnote#1{}
\def\draftmarginnote#1{\marginpar{\raggedright\scriptsize\tt#1}}
\def\draft{
        \pagestyle{plain}
        \overfullrule=2pt
        \oddsidemargin -.5truein
        \def\@oddhead{\sl \phantom{\today\quad\militarytime} \hfil
        \smash{\Large\sl DRAFT} \hfil \today\quad\militarytime}
        \let\@evenhead\@oddhead
        \let\label=\draftlabel
        \let\marginnote=\draftmarginnote
        \def\ps@empty{\let\@mkboth\@gobbletwo
        \def\@oddfoot{\hfil \smash{\Large\sl DRAFT} \hfil}
        \let\@evenfoot\@oddhead}
        \def\@eqnnum{(\theequation)\rlap{\kern\marginparsep\tt\@eqnlabel}%
        \global\let\@eqnlabel\@vacuum}  }

\newcommand{\rf}[1]{(\ref{#1})}
\renewcommand{\theequation}{\thesection.\arabic{equation}}
\renewcommand{\thefootnote}{\fnsymbol{footnote}}
\newcommand{\newsection}{    
\setcounter{equation}{0}\section}

\def\appendix#1{\addtocounter{section}{1}\setcounter{equation}{0}
\renewcommand{\thesection}{\Alph{section}}
\section*{Appendix \thesection\protect\indent \parbox[t]{11.15cm}{#1}}
\addcontentsline{toc}{section}{Appendix \thesection\ \ \ #1}}

\def\nline{\,\nabla\kern -0.7em\raise0.2ex\hbox{/}\,\,}
\def\yline{\,y\kern -0.47em /}
\def\aline{\,a\kern -0.49em /}
\def\parline{\,\partial\kern -0.55em /\,\,}

\def\half{{\frac{1}{2}}}

\newcommand{\Po}{\mathbb{P}}
\newcommand{\Ko}{\mathbb{K}}
\newcommand{\Mo}{\mathbb{M}}
\newcommand{\No}{\mathbb{N}}
\newcommand{\Xo}{\mathbb{X}}

\def\be{\begin{equation}}
\def\ee{\end{equation}}
\def\beq{\begin{eqnarray}}
\def\eeq{\end{eqnarray}}

\def\smone{{\scriptscriptstyle (1)}}
\def\smtwo{{\scriptscriptstyle (2)}}
\def\smthree{{\scriptscriptstyle (3)}}
\def\smfour{{\scriptscriptstyle (4)}}
\def\smfive{{\scriptscriptstyle (5)}}
\def\smsix{{\scriptscriptstyle (6)}}
\def\smseven{{\scriptscriptstyle (7)}}

\def\Ism{{\scriptscriptstyle I}}
\def\IIsm{{\scriptscriptstyle II}}
\def\IIIsm{{\scriptscriptstyle III}}
\def\IVsm{{\scriptscriptstyle IV}}
\def\Vsm{{\scriptscriptstyle V}}

\def\smpt{{\scriptscriptstyle [2]}}
\def\smp3{{\scriptscriptstyle [3]}}

\def\smpn{{\scriptscriptstyle [n]}}

\def\Thsm{{\scriptscriptstyle \rm Th}}

\def\abf{{\bf a}}
\def\bbf{{\bf b}}

\def\Dbf{{\bf D}}
\def\Jbf{{\bf J}}
\def\Kbf{{\bf K}}
\def\kbf{{\bf k}}

\def\Mbf{{\bf M}}
\def\Nbf{{\bf N}}
\def\Pbf{{\bf P}}

\def\Xbf{{\bf X}}

\def\LL{{\cal L}}

\def\Nsf{{\sf N}}

\def\alphab{\bar\alpha}
\def\zetab{\bar\zeta}
\def\upsilonb{\bar\upsilon}

\def\kb{{\bar{k}}}
\def\mb{{\bar{m}}}

\def\add{{\rm add}}
\def\for{{\rm for}}
\def\even{{\rm even}}
\def\ext{{\rm ext}}

\def\dyn{{\rm dyn}}
\def\kin{{\rm kin}}

\def\minrm{{\rm min}}
\def\maxrm{{\rm max}}
\def\irm{{\rm i}}
\def\diff{{\rm diff}}

\def\oplussm{{\scriptscriptstyle \oplus}}
\def\ominussm{{\scriptscriptstyle \ominus}}

\def\opsm{{\scriptscriptstyle \oplus}}
\def\omsm{{\scriptscriptstyle \ominus}}

\def\ommsm{{\scriptscriptstyle \ominus\ominus}}
\def\ompsm{{\scriptscriptstyle \ominus\oplus}}

\def\oppsm{{\scriptscriptstyle \oplus\oplus}}

\def\phik{|\phi\rangle}

\def\Rsm{{\scriptscriptstyle R}}
\def\Lsm{{\scriptscriptstyle L}}

\def\betach{\check{\beta}}

\def\upsilonbf{{\boldsymbol{\upsilon}}}

\def\asf{{\sf a}}
\def\bsf{{\sf b}}
\def\csf{{\sf c}}
\def\esf{{\sf e}}
\def\Nsf{{\sf N}}

\def\noinbf#1{\noindent{\bf #1}}

\begin{document}


\begin{flushright}
FIAN-TD-2016-27 \hspace{1.9cm} {}~\\
arXiv: 1612.06348V2 [hep-th]  {}~ \\

\medskip
Updated September 2022 \hspace{0.8cm} {}~
\end{flushright}

\vspace{1cm}

\begin{center}

{\Large \bf General light-cone gauge approach to conformal fields and
\medskip

applications to scalar and vector fields}

\vspace{2.5cm}

R.R. Metsaev\footnote{ E-mail: metsaev@lpi.ru }

\vspace{1cm}

{\it Department of Theoretical Physics, P.N. Lebedev Physical
Institute, \\ Leninsky prospect 53,  Moscow 119991, Russia }

\vspace{3cm}

{\bf Abstract}

\end{center}

Totally symmetric arbitrary spin conformal fields propagating in the flat space of even dimension greater than or equal to four are studied. For such fields, we develop a general ordinary-derivative light-cone gauge formalism and obtain  restrictions imposed by the conformal algebra symmetries on interaction vertices. We apply our formalism for the detailed study of conformal scalar and vector fields. For such fields, all parity-even cubic interaction vertices are obtained. The cubic vertices obtained are presented in terms of dressing operators and undressed vertices. We show that the undressed vertices of the conformal scalar and vector fields are equal, up to overall factor, to the cubic vertices of massless scalar and vector fields. Various conjectures about interrelations between the cubic vertices for conformal fields in conformal invariant theories and the cubic vertices for massless fields in Poincar\'e invariant theories are proposed.

\vspace{3cm}

\newpage
\renewcommand{\thefootnote}{\arabic{footnote}}
\setcounter{footnote}{0}

\section{Introduction}

Light-cone formulation of relativistic dynamics \cite{Dirac:1949cp}-\cite{Kaku:1974zz}
offers many interesting conceptual and technical simplifications of approaches  to
problems of superstring theories and modern quantum field theory. As the example of a powerful application of light-cone formalism  we can mention the solution to a light-cone gauge superstring field theory \cite{Green:1982tc} and the construction of various supersymmetric theories in terms of off-shell superfields (see, e.g., Refs.\cite{Brink:1982pd}-\cite{Ananth:2005vg}). Light-cone formalism turned also to be helpful for the construction of interaction vertices of higher-spin fields
\cite{Bengtsson:1983pd}-\cite{Metsaev:2007rn} (for recent study of this theme, see Refs.\cite{Bengtsson:2014qza}). Interesting applications of the light-cone formalism for studying gauge/gravity duality may be found in Refs.\cite{Brodsky:2014yha}.
Light-cone gauge formulation of fields dynamics in AdS space and CFT may be found in Refs.\cite{Metsaev:1999ui,Metsaev:2015rda}. In addition to above-said, we also note that, as was demonstrated in  Refs.\cite{Siegel:1986zi}, light-cone gauge formulation may sometimes be a good starting point for deriving new interesting Lorentz covariant formulations.

In this paper, we apply light-cone gauge formalism for study of interaction
vertices of conformal fields. Commonly used formulations of most conformal
fields involve higher derivatives (for review, see
Ref.\cite{Fradkin:1985am}). In Refs.\cite{Metsaev:2007fq,Metsaev:2007rw}, we
developed an ordinary-derivative gauge invariant
formulation for free bosonic conformal fields. Our formulation for free bosonic conformal fields does not involve higher than second order terms in derivatives. Using our ordinary-derivative gauge invariant formulation, we developed the ordinary-derivative light-cone gauge formulation for free bosonic conformal fields in Refs.\cite{Metsaev:2013kaa}. Our ordinary-derivative
light-cone gauge Lagrangian for free bosonic conformal fields in Ref.\cite{Metsaev:2013kaa} does not involve higher than second order terms in derivatives. In this paper, we generalize light-cone gauge formulation of free conformal fields in Ref.\cite{Metsaev:2013kaa} to the case of interacting conformal fields. We develop a method for the building of interaction vertices for arbitrary spin conformal fields and use this method to find explicit expressions for all parity-even cubic vertices for the scalar and vector fields propagating in the $R^{d-1,1}$ space, $d\geq 4$. In this paper, we consider the scalar fields with the conformal dimension $\Delta=\frac{d-2}{2}-k$, $k\in \No_0$, and the vector fields with the conformal dimension $\Delta=1$.

Before proceeding to the main theme in this paper let us briefly review various approaches to conformal fields which have been discussed in the literature.
By using conformal space method, the conformal gravity, for $d=4$, was studied in Ref.\cite{Preitschopf:1998ei}.  By using various methods, Weyl invariant densities of the conformal gravity were obtained, for $d=6$, in Refs.\cite{Bonora:1985cq} and, for $d=8$, in Ref.\cite{Boulanger:2004zf} (see also Ref.\cite{Boulanger:2004eh}). For $d=6$, the ordinary-derivative formulation of the conformal gravity may be found in Ref.\cite{Metsaev:2010kp}. BRST approach to conformal gravity for $d=4$ and arbitrary spin conformal free fields for $d\geq 4$ was discussed in respective Ref.\cite{Boulanger:2001he} and Ref.\cite{Metsaev:2015yyv}. Conformal fields in the AdS space and various curved backgrounds were studied in Refs.\cite{Metsaev:2014iwa}-\cite{Grigoriev:2016bzl}. Mixed-symmetry conformal free fields were investigated in Ref.\cite{Vasiliev:2009ck}.
Scattering amplitudes for conformal fields were studied in Refs.\cite{Beccaria:2016syk}.
Study of interacting higher-spin conformal fields may be found in Refs.\cite{Fradkin:1989md,Segal:2002gd}.

This paper is organized as follows.

In Sec. \ref{seccc-01}, we
describe the ordinary-derivative light-cone gauge formulation of
free bosonic arbitrary spin conformal fields propagating in the $R^{d-1,1}$ space. For $d\geq 4$,
we present $so(d-2)$ covariant version of the formalism. For $d=4$, we also present the helicity
basis formulation of light-cone gauge conformal fields in the $R^{3,1}$ space.

Sec. \ref{seccc-03} is devoted to $n$-point interaction vertices of conformal fields.
We find restrictions imposed on the interaction vertices by kinematical
symmetries of the  conformal $so(d,2)$ algebra.

In Sec. \ref{seccc-04}, we discuss light-cone dynamical principle and present complete list of equations for cubic interaction vertices of arbitrary spin conformal fields. Our equations can be used to find solutions to cubic vertices uniquely up to the freedom related to fields redefinitions.

In Sec. \ref{seccc-05}, we restrict our attention to scalar and vector conformal fields. For such fields, using field redefinitions, we find convenient representatives of parity-even cubic vertices. The equations obtained in Sec. \ref{seccc-04} are reformulated in terms of the representatives of the cubic vertices. The equations obtained can be used to find all solutions to parity-even cubic vertices uniquely.

In Sec.\ref{seccc-06}, for the scalar and vectors fields, we present our solutions for all parity-even cubic vertices in terms of dressing operators and undressed vertices. We show that the dressed vertices obtained coincide, up to some factor, with the cubic vertices for massless and vector fields. For totally symmetric arbitrary spin fields, two conjectures about interrelations between cubic vertices for conformal fields and cubic vertices for massless fields are formulated.

Our conclusions are summarized in Sec.\ref{seccc-07}.

In Appendix A, we prove a statement about representatives of cubic vertices for scalar and vector fields. In Appendix B, we present systematic method for the derivation of cubic vertices of scalar and vector fields in terms of the dressing operators and undressed vertices. In Appendix C, we present the derivation of cubic densities entering dynamical generators of the conformal algebra.

\newsection{Free light-cone gauge conformal fields} \label{seccc-01}

According to the method discussed in Ref.\cite{Dirac:1949cp}, a problem
of finding a new dynamical system  amounts to a problem
of finding a new solution of commutation relations of a basic symmetry algebra.
For the case of conformal fields propagating in the $R^{d-1,1}$ space,
basic symmetries are governed by the conformal algebra $so(d,2)$.
Therefore we start with a discussion of a realization of
the conformal algebra symmetries on a space of conformal fields.
In this section, we focus on free light-cone gauge conformal fields.

\noinbf{Notation and conventions}. The conformal algebra $so(d,2)$ is spanned by the translation generators $P^\mu$, the dilatation generator $D$, the conformal boost generators $K^\mu$ and rotation generators $J^{\mu\nu}$ which are generators of
the $so(d-1,1)$ Lorentz algebra. We use the commutators given by
\beq
&& {}[D,P^\mu]=-P^\mu\,, \hspace{2.5cm}  [P^\mu,J^{\nu\rho}]=\eta^{\mu\nu}P^\rho - \eta^{\mu\rho}P^\nu\,,
\nonumber\\
\label{man-14122016-01} && [D,K^\mu]=K^\mu\,, \hspace{2.7cm} [K^\mu,J^{\nu\rho}]=\eta^{\mu\nu}K^\rho - \eta^{\mu\rho}K^\nu\,,
\\
&& [P^\mu,K^\nu]=\eta^{\mu\nu}D - J^{\mu\nu}\,, \hspace{1.2cm}  [J^{\mu\nu},J^{\rho\sigma}]=\eta^{\nu\rho}J^{\mu\sigma}+3\hbox{ terms} \,,
\nonumber
\eeq
where $\eta^{\mu\nu}$ stands for the mostly positive flat metric tensor.
The translation generators $P^\mu$ and the conformal boost generators $K^\mu$ are chosen to be hermitian, while the dilatation generator $D$ and the Lorentz algebra generators $J^{\mu\nu}$ are considered to be antihermitian. In order to discuss the light-cone gauge approach, we introduce, in place of
the Lorentz basis coordinates $x^\mu$, the light-cone basis
coordinates $x^\pm$, $x^i$ defined by the following relations:%
\footnote{ The indices $\mu,\nu,\rho,\sigma = 0,1,\ldots,d-1$, are vector indices of the $so(d-1,1)$ algebra, while the indices $i,j,k=1,\ldots,d-2$, are vector
indices of the $so(d-2)$ algebra.}
\be \label{man-14122016-04}
x^\pm \equiv \frac{1}{\sqrt{2}}(x^{d-1}  \pm x^0)\,,\qquad
x^i\,,\quad i=1,\ldots, d-2\,.
\ee
Throughout this paper the $x^{+}$ is treated as an evolution parameter. In the light-cone frame, the
Lorentz basis vector $X^\mu$ is decomposed as $X^+,X^-,X^i$, while a
scalar product of two vectors of the Lorentz algebra $so(d-1,1)$ is decomposed as
\be \label{man-14122016-05}
\eta_{\mu\nu} X^\mu Y^\nu = X^+Y^- + X^-Y^+ +X^i Y^i\,.
\ee
Contravariant and covariant components of vectors are
related as $X^+=X_-$, $X^-=X_+$, $X^i=X_i$.

In the light-cone approach, the conformal algebra generators are
separated into the following two groups:
\beq
\label{man-14122016-06} && P^+,\quad P^i,\quad J^{+i},\quad J^{+-},\quad
J^{ij}, \quad D, \ \ K^i,\ \ K^+, \qquad \hbox{ kinematical generators};
\\
\label{man-14122016-07} && P^-,\quad J^{-i}\,, \quad K^-\,, \hspace{6cm} \hbox{ dynamical
generators}.
\eeq
One postulates then that, for $x^+=0$, the kinematical generators \rf{man-14122016-06}  in the field realization are quadratic in the physical fields%
\footnote{ For $x^+ \ne 0$, kinematical generators \rf{man-14122016-06} admit the representation $G= G_1 + x^+ G_2$, where a functional $G_1$ is quadratic in fields, while a functional $G_2$ contains higher order terms in fields.}.
In general, the dynamical generators given in \rf{man-14122016-07}  receive higher-order interaction-dependent corrections.
In the light-cone basis, the commutators of the conformal algebra can be
obtained from the ones in \rf{man-14122016-01} by using the light-cone flat metric having the
following non-vanishing elements: $\eta^{ij}=\delta^{ij}$, $\eta^{+-}=1$, $\eta^{-+}=1$.

To streamline a discussion of the light-cone gauge formulation of conformal fields we use creation operators $\alpha^i$, $\zeta$, $\upsilon^\opsm$, $\upsilon^\omsm$ and the respective annihilation operators $\alphab^i$, $\zetab$, $\upsilonb^\omsm$, $\upsilonb^\opsm$ defined as
\beq
\label{man-14122016-14-a1} && [\alphab^i,\alpha^j]=\delta^{ij}\,,\qquad [\zetab,\zeta] = 1\,,  \qquad [\upsilonb^\opsm,\upsilon^\omsm]=1\,,\qquad [\upsilonb^\omsm,\upsilon^\opsm] = 1\,,
\nonumber\\
&& \alpha^{i \dagger} = \alphab^i \,, \hspace{1.6cm}  \zeta^\dagger = \zetab\,, \hspace{1.4cm}   \upsilon^{\opsm \dagger} = \upsilonb^\opsm \,, \hspace{1.4cm}   \upsilon^{\omsm \dagger} = \upsilonb^\omsm\,,
\nonumber\\
&& \alphab^i |0\rangle = 0\,, \hspace{1.5cm}   \zetab |0\rangle = 0 \,, \hspace{1.1cm} \upsilonb^\oplussm|0\rangle = 0\,, \hspace{1.2cm}  \upsilonb^\omsm|0\rangle = 0 \,.
\eeq
The creation and annihilation operators \rf{man-14122016-14-a1} will be referred to as oscillators in this paper. Here we summarize our notation we use for various quantities constructed out of the oscillators,
\beq
\label{man-14122016-19-a1} && N_\alpha\equiv \alpha^i\alphab^i\,, \qquad N_\zeta\equiv \zeta\zetab\,, \qquad N_{\upsilon^\opsm}\equiv \upsilon^\opsm\upsilonb^\omsm\,, \qquad
N_{\upsilon^\omsm} \equiv \upsilon^\omsm \upsilonb^\opsm\,, \qquad
\nonumber\\
&& \alpha^2 \equiv \alpha^i\alpha^i\,,\qquad \alphab^2 \equiv \alphab^i\alphab^i\,.
\eeq

Denoting arbitrary spin totally symmetric conformal fields by $\phi(x^+,x^-,x^i)$, we assume that these fields are real-valued.
In this paper, we prefer to use Fourier transformed fields $\phi(x^+,p)$,
\beq
\label{man-16092022-01} && \phi(x^+,x^-,x^i) = \int \frac{d^{d-1} p}{(2\pi)^{d-1}} e^{\irm (\beta x^- + x^i p^i)} \phi(x^+,p)\,, \qquad d^{d-1}p\equiv d^{d-2} p\, d\beta\,,
\\
\label{man-16092022-02} && (\phi(x^+,p))^\dagger = \phi(x^+,-p)\,.
\eeq
In \rf{man-16092022-01}, \rf{man-16092022-02}, and below, the argument $p$ in $\phi(x^+,p)$ stands for $p^i,\beta$, where $\beta\equiv p^+$. We now discuss the ordinary-derivative light-cone gauge formulation of various conformal fields in turn.

{\bf Conformal scalar field}. Conformal scalar field  propagating in the $R^{d-1,1}$ space, $d\geq 4$, has conformal dimension $\Delta = \frac{d-2}{2} - k$, $k\in \No_0$. To discuss the ordinary-derivative light-cone gauge formulation of such field  we use $k+1$ scalar fields of the $so(d-2)$ algebra,
\be \label{man-14122016-08}
\phi_{k'}\,,\qquad k' \in [k]_2\,,
\qquad
k \in \No_0\,.
\ee
In \rf{man-14122016-08} and below, the notation $\lambda \in [k]_2$
implies the following convention for the values of $\lambda$:
\be \label{man-14122016-08-a1}
\lambda \in [k]_2 \quad \Longleftrightarrow \quad \lambda
=-k,-k+2,-k+4,\ldots,k-4, k-2,k\,.
\ee
The conformal dimensions of the scalar fields $\phi_{k'}$ \rf{man-14122016-08} are given by
\be \label{man-14122016-09}
\Delta(\phi_{k'}) = \frac{d-2}{2} + k'\,.
\ee
Using the oscillators  $\upsilon^\opsm$, $\upsilon^\omsm$ \rf{man-14122016-14-a1}, we collect all scalar fields \rf{man-14122016-08} into a ket-vector $\phik$ defined as
\be \label{man-14122016-10}
\phik \equiv \sum_{k'\in [k]_2} \frac{1}{ (\frac{k+k'}{2})!}
(\upsilon^\omsm)^{^{\frac{k+k'}{2}}}
(\upsilon^\opsm)^{^{\frac{k-k'}{2}}} \, \phi_{k'} |0\rangle\,.
\ee
The ket-vector $\phik$ \rf{man-14122016-10} is a
degree-$k$ homogeneous polynomial in the oscillators $\upsilon^\oplussm$, $\upsilon^\ominussm$,
\be \label{man-14122016-11sc}
(N_{\upsilon^\opsm} + N_{\upsilon^\omsm}   - k ) \phik = 0 \,.
\ee
Ordinary-derivative Lagrangian for fields \rf{man-14122016-08} may be found in Ref.\cite{Metsaev:2007fq}. In the literature, a conformal scalar field with the conformal dimension $\Delta= \frac{d-2}{2}-k$ with $k>0$, is sometimes  referred to as higher-order singleton. Study of various aspects of the higher-order singleton and related higher-spin algebras may be found, e.g., in Refs.\cite{Bekaert:2013zya,Alkalaev:2014nsa}.

\noindent {\bf Conformal vector field}. Conformal vector field  propagating in the $R^{d-1,1}$ space, $d\geq 4$, has conformal dimension $\Delta = 1$. To discuss the ordinary-derivative light-cone gauge
formulation of such field, we
use $k+1$ vector fields $\phi_{k'}^i$ and $k$ scalar fields $\phi_{k'}$ of the $so(d-2)$ algebra,
\beq
&&\phi_{k'}^i\,, \qquad \ \ \ k' \in [k]_2\,;
\nonumber\\[-16pt]
\label{man-14122016-11} && \hspace{6cm}  k \equiv \frac{d-4}{2}\,, \qquad d-\hbox{even}.
\\[-16pt]
&& \phi_{k'}, \qquad\quad k'\in [k-1]_2\,;
\nonumber
\eeq
The conformal dimensions of the scalar and vector fields \rf{man-14122016-11} are given by
\be \label{man-14122016-14}
\Delta(\phi_{k'}^i)
= \frac{d-2}{2} + k'\,,\qquad
\qquad
\Delta(\phi_{k'})
= \frac{d-2}{2} + k'\,.
\ee
We note that the scalar fields $\phi_{k'}$ \rf{man-14122016-11} enter the field content only when $d \geq 6$, i.e., $k\geq 1$.

Using the oscillators  $\alpha^i$, $\zeta$, $\upsilon^\opsm$, $\upsilon^\omsm$, we collect fields \rf{man-14122016-11} into a ket-vector $\phik$ defined as
\beq
\label{man-14122016-15} &&  \phik =  |\phi_1 \rangle  + \zeta |\phi_0\rangle
\,,
\\
\label{man-14122016-16} && |\phi_1\rangle \equiv \sum_{k'\in [k]_2}  \frac{1}{
(\frac{k+k'}{2})!} \alpha^i (\upsilon^\omsm)^{^{\frac{k+k'}{2}}}(\upsilon^\opsm)^{^{\frac{k-k'}{2}}}
 \, \phi_{k'}^i  |0\rangle\,,
\\
\label{man-14122016-17} && |\phi_0\rangle \equiv \sum_{k'\in [k-1]_2} \frac{1}{
(\frac{k-1+k'}{2})!} (\upsilon^\omsm)^{^{\frac{k-1+k'}{2}}}
(\upsilon^\opsm)^{^{\frac{k-1-k'}{2}}}   \, \phi_{k'}^{\vphantom{5pt}}
|0\rangle\,.
\eeq
Using \rf{man-14122016-15}-\rf{man-14122016-17}, we verify that $\phik$, $|\phi_1\rangle$, and $|\phi_0\rangle$ obey the following constraints:
\beq
\label{man-14122016-18}  && (N_\alpha + N_\zeta)\phik =  \phik\,,\hspace{3cm}
(N_\zeta + N_{\upsilon^\opsm} + N_{\upsilon^\omsm})\phik = k \phik\,,
\\
\label{man-14122016-19} && (N_{\upsilon^\opsm} + N_{\upsilon^\omsm}) |\phi_1\rangle = k |\phi_1\rangle\,,\hspace{2cm} (N_{\upsilon^\opsm} + N_{\upsilon^\omsm}) |\phi_0\rangle = (k-1) |\phi_0\rangle\,.\qquad
\eeq
From algebraic constraints \rf{man-14122016-18}, we learn that the ket-vector $\phik$ is
degree-1 homogeneous polynomial in the oscillators $\alpha^i$, $\zeta$,  and
degree-$k$ homogeneous polynomial in the oscillators $\zeta$,
$\upsilon^\oplussm$, $\upsilon^\ominussm$. From algebraic constraints \rf{man-14122016-19}, we learn that the ket-vectors $|\phi_1\rangle$ and $|\phi_0\rangle$ are the respective
degree-$k$ and $k-1$ homogeneous polynomials in the oscillators
$\upsilon^\oplussm$, $\upsilon^\ominussm$.

\noindent {\bf Arbitrary spin-$s$ conformal field}. Totally symmetric spin-$s$ conformal field propagating in the $R^{d-1,1}$ space, $d\geq 4$, has conformal dimension $\Delta = 2-s$. To develop the ordinary-derivative light-cone gauge formulation for such field we use the following fields of the $so(d-2)$ algebra:
\beq
\label{man-14122016-20} && \phi_{k'}^{i_1\ldots i_{s'}}\,,
\hspace{2cm}
s'= \left\{\begin{array}{l}
0,1,\ldots,s;\qquad  \hbox{for }\  d \geq 6;
\\[3pt]
1,2,\ldots,s;\qquad  \hbox{for }d = 4;
\end{array}\right.
\hspace{2cm} k' \in  [k_{s'}]_2\,;\qquad
\\
\label{man-14122016-20-a1} && \Delta(\phi_{k'}^{i_1\ldots i_{s'}}) = \frac{d-2}{2} + k'\,,\qquad \hspace{2cm} k_{s'} \equiv s'+ \frac{d-6}{2}\,, \qquad d-\hbox{even}\,,
\eeq
where fields $\phi_{k'}$ and $\phi_{k'}^i$ are the respective scalar and vector fields of the $so(d-2)$ algebra, while field
$\phi_{k'}^{i_1\ldots i_{s'}}$, $s'\geq 2$, is rank-$s'$ totally
symmetric traceless tensor field of the $so(d-2)$ algebra,
\be \label{man-14122016-21}
\phi_{k'}^{iii_3\ldots i_{s'}}=0\,, \qquad
s'\geq 2\,.
\ee
From \rf{man-14122016-20}, we see that the scalar fields $\phi_{k'}$ enter the field content only when $d\geq 6$.
Alternatively, the field content \rf{man-14122016-20}
can be represented as
\beq
\label{man-14122016-22} &&\phi_{k'}^{i_1\ldots i_s}\,, \hspace{2.5cm} k'\in [k_s]_2\,;
\\
\label{man-14122016-23} &&\phi_{k'}^{i_1\ldots i_{s-1}}\,, \hspace{2.2cm} k'\in [k_s-1]_2\,;
\\
&&\ldots \quad \ldots \hspace{2.5cm}  \ldots \quad \ldots
\nonumber\\
&&\ldots \quad \ldots \hspace{2.5cm} \ldots \quad \ldots
\nonumber\\
\label{man-14122016-24} && \phi_{k'}^i\,, \hspace{3cm} k'\in
[k_s-s+1]_2\,;
\\
\label{man-14122016-25} && \phi_{k'}\,, \hspace{3cm} k'\in [k_s-s]_2\,;
\eeq
\be \label{man-14122016-26}
k_s \equiv s+ \frac{d-6}{2}\,.
\ee
Namely, for $d\geq 6$, the field content is given in \rf{man-14122016-22}-\rf{man-14122016-25},
while, for $d=4$, the field content is given in \rf{man-14122016-22}-\rf{man-14122016-24}.

Using the oscillators \rf{man-14122016-14-a1}, we introduce the ket-vector $\phik$,%
\footnote{ Among other things, we use the $so(d-2)$ algebra vector oscillators $\alpha^i$. Use of twistor-like variables for a discussion of conformal fields may be found in Refs.\cite{Uvarov:2014lfa,Adamo:2016ple}. We note also that Lorentz algebra vector oscillators are popular in the framework of BRST approach to higher-spin fields (see, e.g., Refs.\cite{Buchbinder:2011xw})
}
\beq
&& \phik \equiv \sum_{s'=0}^s
\frac{\zeta^{s-s'}}{\sqrt{(s-s')!}}|\phi_{s'}\rangle \,,
\hspace{3cm} \hbox{for } \ d\geq 6\,,
\nonumber\\[-10pt]
&& \label{man-14122016-27}
\\[-10pt]
&& \phik \equiv \sum_{s'=1}^s
\frac{\zeta^{s-s'}}{\sqrt{(s-s')!}}|\phi_{s'}\rangle \,,
\hspace{3cm} \hbox{for } \ d =4 \,,
\nonumber\\
\label{man-14122016-28} && |\phi_{s'}\rangle \equiv   \sum_{k'\in
[k_{s'}]_2} \frac{1}{s'!(\frac{k_{s'} + k'}{2})!}\alpha^{i_1} \ldots
\alpha^{i_{s'}} (\upsilon^\omsm)^{^{\frac{k_{s'}+k'}{2}}}
(\upsilon^\opsm)^{^{\frac{k_{s'} - k'}{2}}} \,
\phi_{k'}^{i_1\ldots i_{s'}}|0\rangle\,.
\eeq
From \rf{man-14122016-27}, \rf{man-14122016-28},  we learn that the ket-vectors $\phik$, $|\phi_{s'}\rangle$  satisfy the following restrictions
\beq
\label{man-14122016-29} && (N_\alpha + N_\zeta - s)\phik = 0 \,,
\hspace{2cm} (N_\zeta + N_{\upsilon^\opsm} + N_{\upsilon^\omsm} - k_s) \phik = 0 \,,
\\
\label{man-14122016-30} && (N_\alpha - s') |\phi_{s'}\rangle = 0
\,, \hspace{2.6cm}  (N_{\upsilon^\opsm} + N_{\upsilon^\omsm} - k_{s'}) |\phi_{s'}\rangle =0 \,,
\eeq
where $k_{s'}$ is given in \rf{man-14122016-20-a1}. From algebraic constraints
\rf{man-14122016-29}, we learn that the ket-vector $\phik$ is degree-$s$ homogeneous polynomial in the oscillators $\alpha^i$, $\zeta$ and degree-$k_s$ homogeneous polynomial in the oscillators
$\zeta$, $\upsilon^\oplussm$, $\upsilon^\ominussm$, while, from algebraic constraints
\rf{man-14122016-30}, we learn that the ket-vector
$|\phi_{s'}\rangle$ is degree-$s'$ homogeneous polynomial in the
oscillators $\alpha^i$ and degree-$k_{s'}$ homogeneous polynomial in
the oscillators $\upsilon^\oplussm$, $\upsilon^\ominussm$. In terms of $\phik$, the constraint \rf{man-14122016-21} takes the form $\alphab^2 \phik = 0$.

We now describe a realization of the conformal algebra symmetries on a space of conformal fields. Let us use the notation $|\phi_{\Delta,s}(x^+,p,\alpha)\rangle$ for the ket-vectors given in \rf{man-14122016-10}, \rf{man-14122016-15}, \rf{man-14122016-27}, where a shortcut $\alpha$ stands for the oscillators $\alpha^i$, $\zeta$, $\upsilon^\opsm$, $\upsilon^\omsm$, while the subscripts $\Delta$, $s$ indicate that we deal with spin-$s$ conformal field having conformal dimension $\Delta$. A representation of the kinematical and dynamical generators in terms of differential operators acting on the ket-vector $|\phi_{\Delta,s}(x^+,p,\alpha)\rangle$ is given by

\noinbf{kinematical generators}:
\beq
\label{man-14122016-32} && P^i=p^i\,, \hspace{0.5cm} P^+=\beta\,,
\\
\label{man-14122016-34} && J^{+-} = {\rm i}x^+P^- + \partial_\beta \beta\,, \hspace{1cm} J^{+i}= {\rm i} x^+ p^i + \partial_{p^i}\beta\,,
\\
\label{man-14122016-36} && J^{ij}= p^i\partial_{p^j} - p^j\partial_{p^i}+M^{ij}\,,
\\
\label{man-14122016-33} && D = {\rm i}x^+ P^- -\partial_\beta \beta - \partial_{p^i}p^i + \frac{d-2}{2} + M^\ompsm\,,
\\
\label{man-14122016-37} && K^+ = \frac{1}{2}(2{\rm i} x^+ \partial_\beta - \partial_{p^i}\partial_{p^i} +  M^\ommsm)\beta +  {\rm i} x^+ D\,,
\\
\label{man-14122016-38} && K^i = \frac{1}{2}(2{\rm i} x^+ \partial_\beta - \partial_{p^j} \partial_{p^j} +  M^\ommsm)p^i
-\partial_{p^i} D - M^{ij}\partial_{p^j} + M^{\ominussm i}  + {\rm i} M^{i-}x^+\,, \qquad
\eeq
\noinbf{dynamical generators}:
\beq
\label{man-14122016-39} && P^- = - \frac{p^i p^i + M^\oppsm}{2\beta}\,,
\\
\label{man-14122016-40} && J^{-i} = -\partial_\beta p^i + \partial_{p^i} P^- + M^{-i}\,,
\\
\label{man-14122016-41} && K^- =\half (2\irm x^+ \partial_\beta - \partial_{p^i}\partial_{p^i} + M^\ommsm)P^- -\partial_\beta D - \partial_{p^i} M^{-i} - M^{\ominussm i} \frac{p^i}{\beta} +  \frac{1}{\beta}B   \,,\qquad
\eeq
where we use the following notation:
\beq
\label{man-14122016-42} &&  M^\ompsm =  \upsilon^\omsm \upsilonb^\opsm - \upsilon^\opsm \upsilonb^\omsm \,, \hspace{1cm} M^\oppsm = \upsilon^\oplussm \upsilonb^\oplussm\,, \hspace{0.9cm}  M^\ommsm = 4 \upsilon^\ominussm \upsilonb^\ominussm \,, \qquad
\\
\label{man-14122016-43} && M^{ij}=  \alpha^i \alphab^j - \alpha^j \alphab^i\,,
\nonumber\\
\label{man-14122016-44} && M^{\opsm i} = m^\opsm \alphab^i + \alpha_\Thsm^i \mb^\opsm\,, \hspace{1cm} M^{\omsm i} = m^\omsm\alphab^i + \alpha_\Thsm^i \mb^\omsm\,,
\\
\label{man-14122016-45} && M^{-i} = M^{ij}\frac{p^j}{\beta} + \frac{1}{\beta}  M^{\oplussm i} \,,
\\
\label{man-14122016-46} && B  = - s - N_\zeta (2s+d-4-N_\zeta)\,,
\\
\label{man-14122016-47} && \alpha_\Thsm^i = \alpha^i - \alpha^2 \frac{1}{2N_\alpha + d-2}\alphab^i\,,
\\
\label{man-14122016-48} && m^\opsm = \zeta e_\zeta \upsilonb^\oplussm\,, \hspace{1.4cm} \mb^\opsm = - \upsilon^\oplussm e_\zeta \zetab\,,
\nonumber\\
&& m^\omsm = - 2 \zeta e_\zeta \upsilonb^\ominussm\,, \qquad \mb^\omsm = - 2 \upsilon^\ominussm e_\zeta \zetab\,, \hspace{1cm} e_\zeta = \Bigl(\frac{2s+d-4-N_\zeta}{2s+d-4-2N_\zeta}\Bigr)^{1/2}\,.\qquad
\eeq
For the notation, see also \rf{man-14122016-19-a1} and
\be \label{man-14122016-53}
\beta\equiv p^+\,,\qquad
\partial_\beta\equiv \partial/\partial \beta\,,
\quad
\partial_{p^i}\equiv \partial/\partial p^i\,.
\ee
For the reader convenience, we note the following algebra of relations for operators $M^\oppsm$, $M^\ommsm$, $M^\ompsm$, $M^{ij}$, $M^{\opsm i}$, $M^{\omsm i}$,
\beq
\label{man-14122016-54} && [M^\ommsm ,M^\oppsm ]=4M^\ompsm \,,
\nonumber\\
\label{man-14122016-55} && [M^\ompsm ,M^\oppsm]= -2M^\oppsm\,,
\nonumber\\
\label{man-14122016-56} && [M^\ompsm ,M^\ommsm]=   2 M^\ommsm\,,
\\[5pt]
\label{man-14122016-57} && [M^{ij},M^{kl}] = \delta^{jk}M^{il} + 3\hbox{ terms},
\\[5pt]
\label{man-14122016-58} && [M^\ompsm , M^{\opsm i}]= -M^{\opsm i}\,,
\nonumber\\
\label{man-14122016-59} && [M^\ompsm ,M^{\omsm i}]=M^{\omsm i}\,,
\nonumber\\
\label{man-14122016-60} && [M^\oppsm, M^{\omsm i}] = 2 M^{\opsm i}\,,
\nonumber\\
\label{man-14122016-61} && [M^\ommsm, M^{\opsm i}] = 2 M^{\omsm i}\,,
\\[5pt]
\label{man-14122016-62} && [M^{ij},M^{\opsm k}] = \delta^{jk} M^{\opsm i} - \delta^{ik} M^{\opsm j} ,
\nonumber\\
\label{man-14122016-63} && [M^{ij},M^{\omsm k}] = \delta^{jk} M^{\omsm i} - \delta^{ik} M^{\omsm j} ,
\\[5pt]
\label{man-14122016-64} && [M^{\oplussm i},M^{\oplussm j}] = -M^\oppsm M^{ij}\,,
\nonumber\\
\label{man-14122016-65} && [M^{\ominussm i},M^{\ominussm j}] =  M^\ommsm M^{ij}\,,
\\
\label{man-14122016-66} && -\frac{1}{2}\{M^{il},M^{lj}\} + [M^{\opsm i},M^{\omsm j}] + M^{ij}M^\ompsm  = B\delta^{ij}\,,
\eeq
where $\{x,y\}= xy+yx$,  $[x,y]  = xy - yx$.

For scalar and vector conformal fields, the operators $M^{\opsm i}$, $M^{\omsm i}$ given in \rf{man-14122016-44} take the form,
\beq
\label{man-14122016-67} && \hspace{-2cm} \hbox{\it scalar field:} \quad  M^{\opsm i} = 0\,, \quad  M^{\omsm i} = 0\,,  \quad m^\opsm=0\,,  \quad m^\omsm=0\,,  \quad \mb^\opsm=0\,,  \quad \mb^\omsm=0\,;
\nonumber\\
&& \hspace{-2cm} \hbox{\it vector field:} \quad M^{\opsm i} = m^\opsm \alphab^i + \alpha^i \mb^\opsm\,, \qquad M^{\omsm i} = m^\omsm \alphab^i + \alpha^i \mb^\omsm\,,
\nonumber\\
&& \hspace{1.3cm} m^\opsm = \zeta \upsilonb^\opsm\,, \hspace{2.8cm} m^\omsm = -2 \zeta \upsilonb^\omsm\,,
\nonumber\\
&& \hspace{1.3cm}  \mb^\opsm = - \upsilon^\opsm \zetab \,, \hspace{2.5cm} \mb^\omsm = -2  \upsilon^\omsm \zetab\,.
\eeq

Expressions given in \rf{man-14122016-32}-\rf{man-14122016-41} provide the realization of the conformal algebra $so(d,2)$
in terms of  differential operators acting on the physical field
$|\phi\rangle$. To discuss light-cone gauge interaction vertices we use a field theoretical realization of the conformal algebra in terms of the physical field $|\phi\rangle$. To quadratic order in the field $\phik$, the kinematical generators $G^\kin$ and the dynamical generators $G^\dyn$ can be presented as
\be \label{man-15122016-01}
G_\smpt=\int \beta d^{d-1}p\, \langle\phi| G_\diff \phik \,, \qquad \phik\equiv |\phi_{\Delta,s}(x^+,p,\alpha)\rangle\,,
\qquad d^{d-1}p \equiv d\beta d^{d-2}p\,,
\ee
where $G_\diff$ in \rf{man-15122016-01} are the conformal algebra generators
realized as the differential operators given in \rf{man-14122016-32}-\rf{man-14122016-41}.
The bra-vectors are defined as $\langle\phi| \equiv \phik{}^\dagger$. The field $|\phi\rangle$ satisfies the Poisson-Dirac commutator,
\be \label{man-15122016-02}
[\,|\phi_{\Delta,s}(x^+,p,\alpha)\rangle\,,\,|\phi_{\Delta',s'}(x^+,p^\prime,\alpha')\rangle\,] =  \frac{1}{2\beta} \delta(\beta+\beta^\prime)\delta^{d-2}(p+p^\prime)
|\rangle |\rangle' \delta_{\Delta,\Delta'} \delta_{s,s'}\,,
\ee
where $|\rangle |\rangle'$ stands for a projector that respects the algebraic constraints \rf{man-14122016-21}, \rf{man-14122016-29}, \rf{man-14122016-30}.  Using these relations, we get the standard commutator,
\be \label{man-15122016-03}
[ |\phi_{\Delta,s}\rangle, G_\smpt\,] = G_\diff|\phi_{\Delta,s}\rangle\,.
\ee
As noted in the literature (see, e.g., Refs.\cite{Green:1982tc}), in the framework of light-cone gauge approach, a Lagrangian for light-cone gauge interacting fields takes the standard form,
\be \label{man-15122016-04}
S
=
\int dx^+ d^{d-1} p\,\, \langle \phi| \irm \beta
\partial^- \phik + \int dx^+ P^-\,,
\ee
where $P^-$ is the light-cone Hamiltonian.
Incorporation of a internal symmetry into theory of interacting conformal fields can be done via the Chan--Paton method used in string theory \cite{Paton:1969je} (see at the end of Sec.\ref{seccc-06} in this paper).

\noinbf{Helicity basis for light-cone gauge conformal fields}. To discuss conformal fields in the $R^{3,1}$ space, we can use a helicity basis. Though we do not use such basis in this paper we decided, for the reader convenience, to outline briefly the helicity basis formulation of light-cone gauge conformal fields. To this end we introduce a frame of complex coordinates $x^\Rsm$, $x^\Lsm$ defined by the relations
\be \label{man-15122016-04-b1}
x^\Rsm \equiv \frac{1}{\sqrt{2}} (x^1 + \irm x^2)\,, \qquad
x^\Lsm \equiv \frac{1}{\sqrt{2}} (x^1 - \irm x^2)\,.
\ee
In such frame, a vector of the $so(2)$ algebra $X^i$ is decomposed as $X^i = X^\Rsm, X^\Lsm$, while a scalar product of two vectors $X^i$, $Y^i$ is represented as $X^i Y^i = X^\Rsm Y^\Lsm + X^\Lsm Y^\Rsm$.
In the helicity basis, we decompose the oscillators as $\alpha^i=\alpha^\Rsm,\alpha^\Lsm$,
and, in place of the ket-vector $\phik$ given in \rf{man-14122016-27}, \rf{man-14122016-28}, we use a ket-vector defined as
\beq
\label{man-15122016-04-b2} && \phik \equiv \sum_{s'=1}^s \frac{\zeta^{s-s'}}{\sqrt{(s-s')!}}|\phi_{s'}\rangle \,,
\nonumber\\
\label{man-15122016-04-b3} && |\phi_{s'}\rangle \equiv   \sum_{k'\in
[s'-1]_2} \frac{1}{\sqrt{s'!} \, (\frac{s'-1 + k'}{2})!} (\alpha^\Rsm)^{s'} (\upsilon^\omsm)^{^{\frac{s'-1+k'}{2}}}
(\upsilon^\opsm)^{^{\frac{s'-1 - k'}{2}}} \,
\phi_{k',s'}|0\rangle\,,
\eeq
where $\phi_{k',s'}=\phi_{k',s'}(x^+,p)$ are complex-valued fields with the positive helicities $s'$, $s'=1,2,\ldots,s$, $s>0$, and the conformal dimension $k'+1$. Note also the following hermitian conjugation rule for the helicity basis conformal fields: $(\phi_{k',s'}(x^+,p))^\dagger=\phi_{k',-s'}(x^+,-p)$.

We note also that, in the helicity basis, the operators $M^{\opsm i}$, $M^{\omsm i}$ are decomposed as $M^{\opsm i} = M^{\opsm \Rsm}, M^{\opsm\Lsm}$, $M^{\omsm i} = M^{\omsm \Rsm}, M^{\omsm\Lsm}$, while the $so(2)$ algebra generator $M^{ij}=-M^{ji}$ is represented as $M^{\Rsm\Lsm}$. A realization of the conformal algebra on ket-vector $\phik$ \rf{man-15122016-04-b2} can easily be obtained by plugging into \rf{man-14122016-32}-\rf{man-14122016-41} the following expressions for the various operators:
\beq
\label{man-220922022-01} &&  M^\ompsm =  \upsilon^\omsm \upsilonb^\opsm - \upsilon^\opsm \upsilonb^\omsm \,, \hspace{1cm} M^\oppsm = \upsilon^\oplussm \upsilonb^\oplussm\,, \hspace{0.9cm}  M^\ommsm = 4 \upsilon^\ominussm \upsilonb^\ominussm \,, \qquad
\\
&& M^{\Rsm\Lsm} = \alpha^\Rsm \alphab^\Lsm\,,
\nonumber\\
&& M^{\oplussm \Rsm} = \alpha^\Rsm \mb^\opsm\,, \hspace{1cm} M^{\oplussm \Lsm} =  m^\opsm \alphab^\Lsm\,,
\nonumber\\
&& M^{\ominussm\Rsm} =  \alpha^\Rsm \mb^\omsm\,, \hspace{1cm}  M^{\ominussm\Lsm} =   m^\omsm \alphab^\Lsm\,,
\\
&& M^{-\Rsm} = M^{\Rsm\Lsm}\frac{p^\Rsm}{\beta} + \frac{1}{\beta}  M^{\oplussm \Rsm} \,, \hspace{1cm} M^{-\Lsm} = - M^{\Rsm\Lsm}\frac{p^\Lsm}{\beta} + \frac{1}{\beta}  M^{\oplussm \Lsm} \,,
\\
&& B  = - s - N_\zeta (2s-N_\zeta)\,,
\\
&& m^\opsm = \zeta e_\zeta \upsilonb^\oplussm\,, \hspace{1.4cm} \mb^\opsm = - \upsilon^\oplussm e_\zeta \zetab
\nonumber\\
\label{man-220922022-10} && m^\omsm = - 2 \zeta e_\zeta \upsilonb^\ominussm\,, \qquad \mb^\omsm = - 2 \upsilon^\ominussm e_\zeta \zetab\,, \hspace{1cm} e_\zeta = \Bigl(\frac{2s-N_\zeta}{2s-2N_\zeta}\Bigr)^{1/2}\,,
\\
\label{man-15122016-04-b18} && \hspace{0.2cm} [\alphab^\Lsm,\alpha^\Rsm] = 1\,, \qquad[\alphab^\Rsm,\alpha^\Lsm] = 1\,, \qquad   \alpha^{\Rsm \dagger} =\alphab^\Lsm \,,\qquad   \alpha^{\Lsm \dagger} =\alphab^\Rsm \,,
\eeq
where, in \rf{man-15122016-04-b18}, we present commutators of the oscillators and their hermitian properties. To quadratic order in the ket-vector $\phik$ \rf{man-15122016-04-b2}, a field theoretical representation for generators of the $so(4,2)$ algebra takes the form
\be \label{man-15122016-01-b19}
G_\smpt = 2\int \beta d^3p\, \langle\phi| G_\diff \phik\,,\qquad  \qquad \phik\equiv |\phi_{\Delta,s}(x^+,p,\alpha)\,, \qquad \langle\phi|   \equiv \phik{}^\dagger\,,
\ee
$s>0$, $\Delta=2-s$, where $G_\diff$ in \rf{man-15122016-01-b19} are obtained by using $G_\diff$ \rf{man-14122016-32}-\rf{man-14122016-41} and \rf{man-220922022-01}-\rf{man-220922022-10}.
The Poisson-Dirac commutator for the component fields entering ket-vector \rf{man-15122016-04-b2} takes the form
\be
[\phi_{k',s'}(x^+,p')\,, \phi_{k'',s''}(x^+,p'')^\dagger]
=  \frac{1}{2\beta'} \delta(\beta'-\beta'')\delta^2(p'-p'') \delta_{k',k''} \delta_{s',s''}\,.
\ee

\newsection{\large
Kinematical equations for $n$-point interaction vertices} \label{seccc-03}

We now discuss a general structure of the conformal algebra dynamical generators \rf{man-14122016-07}. Recall that, in general, in theories of interacting conformal fields, the dynamical generators receive corrections involving higher powers of physical fields. The dynamical generators are expanded as
\be \label{man-15122016-05}
G^\dyn = \sum_{n=2}^\infty G_\smpn^\dyn\,,
\ee
where $G_\smpn^\dyn$ appearing in \rf{man-15122016-05} stands for the functional
that has $n$ powers of physical fields $\phik$.

Restrictions on $G_\smpn^\dyn$, $n\geq 3$, that are obtained from the
commutators between kinematical generators \rf{man-14122016-06} and dynamical generators \rf{man-14122016-07} we refer to as kinematical symmetry restrictions or kinematical equations,  while restrictions on $G_\smpn^\dyn$, $n\geq 3$, that are obtained from
commutators between the dynamical generators \rf{man-14122016-07} are referred to as dynamical equations. Our aim in this section is to discuss the general structure of $G_\smpn^\dyn$, $n\geq 3$ implied by the kinematical equations. We now discuss the kinematical symmetry restrictions and the kinematical equations.

\noinbf{$P^i$-, $P^+$-kinematical symmetry restrictions}. Using the commutators  between the dynamical generators \rf{man-14122016-07} and the kinematical generators $P^i$ and $P^+$, we learn that the dynamical generators $G_\smpn^\dyn$, $n\geq 3$, can be presented as
\beq
 \label{man-15122016-06} && P_\smpn^- = \int\!\! d\Gamma_n \,\, \langle \Phi_\smpn ||p_\smpn^-\rangle \,,
\\
\label{man-15122016-07} && J_\smpn^{-i} = \int \!\! d\Gamma_n\,\, \langle \Phi_\smpn | j_\smpn^{-i}\rangle + (\Xbf^i \langle \Phi_\smpn |)|p_\smpn^-\rangle \,,
\\
\label{man-15122016-08} && K_\smpn^- =\int\!\! d\Gamma_n\,\, \langle \Phi_\smpn |k_\smpn^-\rangle - ( \Xbf^i \langle \Phi_\smpn|) |j_\smpn^{-i}\rangle - \half (\Xbf^i \Xbf^i \langle \Phi_\smpn| )|p_\smpn^-\rangle  \,,
\eeq
where we use the notation
\beq
\label{man-15122016-09} && \langle \Phi_\smpn| \equiv \prod_{a=1}^n \langle \phi_{\Delta_a,s_a}(x^+,p_a,\alpha_a)|\,,\qquad\qquad
\\
\label{man-15122016-10} && d\Gamma_n \equiv (2\pi)^{d-1} \delta^{d-1}(\sum_{a=1}^np_a)
\prod_{a=1}^n \frac{d^{d-1} p_a}{(2\pi)^{(d-1)/2}} \,,
\\
\label{man-15122016-11} && \Xbf^i \equiv - \frac{1}{n}\sum_{a=1}^n \partial_{p_a^i}\,,
\eeq
while the ket-vectors of densities $|p_\smpn^-\rangle$, $|j_\smpn^{-i}\rangle$, and $|k_\smpn^-\rangle$  \rf{man-15122016-06}-\rf{man-15122016-08} can be presented as
\beq
\label{man-15122016-14} && |p_\smpn^-\rangle = p_\smpn^- (p_a,\beta_a;\, \alpha_a)|0\rangle\,,
\nonumber\\
&&  |j_\smpn^{-i}\rangle = j_\smpn^{-i} (p_a,\beta_a;\, \alpha_a)|0\rangle\,,  \qquad |k_\smpn^-\rangle = k_\smpn^- (p_a,\beta_a;\, \alpha_a)|0\rangle\,.
\eeq
In \rf{man-15122016-09}-\rf{man-15122016-14}, the indices $a,b=1,\ldots,n$ label $n$ interacting
fields. The Dirac $\delta$- function appearing in \rf{man-15122016-10}
implies conservation laws for the momenta $p_a^i$ and $\beta_a$. The densities $p_\smpn^-$, $j_\smpn^{-i}$, and  $k_\smpn^-$
\rf{man-15122016-14} depend on the momenta $p_a^i$, $\beta_a$, and spin variables
denoted by a shortcut $\alpha_a$. The shortcut $\alpha_a$ stands for the oscillators $\alpha_a^i$, $\zeta_a$, $\upsilon_a^\opsm$, $\upsilon_a^\omsm$. The density $p_\smpn^-$ will be referred to as an $n$-point interaction vertex (or cubic interaction vertex when $n=3$).
Sometimes we use a shortcut $(\Delta_a,s_a)$ for the bra-vector $\langle\phi_{\Delta_a,s_a}(x^+,p,\alpha)|$\,,

\be \label{man-20092022-01}
(\Delta_a,s_a) \hspace{1cm} \hbox{ spin-$s_a$ and conformal dimension-$\Delta_a$ field}.
\ee

\medskip
\noinbf{$J^{+-}$-kinematical equations}. Commutators between the dynamical generators \rf{man-14122016-07} and the kinematical generator $J^{+-}$ give the equations%
\footnote{ From now on, in place of equations for the ket-vectors $|g_\smpn\rangle$, we use equations for the densities $g_\smpn$ \rf{man-15122016-14}. To this end, in equations for $g_\smpn$, we replace the annihilation operators $\alphab^i$, $\zetab$, $\upsilonb^\opsm$, and $\upsilonb^\omsm$ \rf{man-14122016-14-a1} by the respective derivatives $\partial / \partial \alpha^i$, $\partial / \partial \zeta$, $\partial / \partial \upsilon^\omsm$, and $\partial / \partial \upsilon^\opsm$.
}

\be \label{man-15122016-17}
\sum_{a=1}^n \beta_a\partial_{\beta_a} \,  p_\smpn^- = 0 \,,\qquad  \sum_{a=1}^n \beta_a\partial_{\beta_a} \,  j_\smpn^{-i} = 0 \,, \qquad  \sum_{a=1}^n \beta_a\partial_{\beta_a} \,  k_\smpn^- = 0 \,.
\ee

\noinbf{$D$-kinematical equations}. Commutators between the dynamical generators \rf{man-14122016-07} and the kinematical generator $D$ give the equations
\beq
\label{man-15122016-20} && \sum_{a=1}^n  p_a^i\partial_{p_a^i}\, p_\smpn^- = \Bigl(2+ \frac{d-2}{2}(2-n)  + \sum_{a=1}^n M_a^{\ompsm} \Bigr) p_\smpn^-\,,
\\
\label{man-15122016-21} && \sum_{a=1}^n  p_a^i\partial_{p_a^i}\, j_\smpn^{-i} = \Bigl(1  + \frac{d-2}{2}(2-n)  + \sum_{a=1}^n M_a^{\ompsm} \Bigr) j_\smpn^{-i}\,,
\\
\label{man-15122016-22} && \sum_{a=1}^n  p_a^i\partial_{p_a^i}\,  k_\smpn^- = \Bigl(\frac{d-2}{2} (2-n)  + \sum_{a=1}^n M_a^{\ompsm} \Bigr) k_\smpn^-\,.
\eeq

\noinbf{$J^{ij}$-kinematical equations}. Commutators between the dynamical generators \rf{man-14122016-07} and the kinematical generators $J^{ij}$ lead to  the equations
\beq
\label{man-15122016-23} && \sum_{a=1}^n (p_a^i\partial_{p_a^j} - p_a^j\partial_{p_a^i}  + M_a^{ij}  )  p_\smpn^- =  0 \,,
\\
\label{man-15122016-24} && \sum_{a=1}^n (p_a^i\partial_{p_a^j} - p_a^j\partial_{p_a^i}  + M_a^{ij}  ) j_\smpn^{-k} = \delta^{jk} j_\smpn^{-i} - \delta^{ik} j_\smpn^{-j}\,,
\\
\label{man-15122016-25} && \sum_{a=1}^n (p_a^i\partial_{p_a^j} - p_a^j\partial_{p_a^i}  + M_a^{ij}  ) k_\smpn^- = 0\,.
\eeq

\noinbf{$K^+$-kinematical equations}. Commutators between the dynamical generators \rf{man-14122016-07} and the kinematical generator $K^+$ lead to the equations
\beq
\label{man-15122016-36} && \sum_{a=1}^n \beta_a \big(M_a^\ommsm - \partial_{p_a^j}\partial_{p_a^j} \big)  p_\smpn^- =  0 \,,
\\
\label{man-15122016-37} && \sum_{a=1}^n \beta_a \big(M_a^\ommsm - \partial_{p_a^j}\partial_{p_a^j} \big)  j_\smpn^{-i} =  0 \,,
\\
\label{man-15122016-38} && \sum_{a=1}^n \beta_a \big(M_a^\ommsm - \partial_{p_a^j}\partial_{p_a^j} \big)  k_\smpn^- =  0 \,.
\eeq

\noinbf{$K^i$-kinematical equations}. Commutators between the dynamical generators \rf{man-14122016-07} and the kinematical generators $K^i$ give the equations
\beq
\label{man-15122016-26} &&    j_\smpn^{-i} -  \Kbf^i p_\smpn^- -  \Xbf^i  p_\smpn^-  = 0 \,,
\\
\label{man-15122016-27} &&  \Kbf^i j_\smpn^{-j} - \Xbf^j j_\smpn^{-i}    + \delta^{ij} \Xbf^l j_\smpn^{-l}
 -  \Kbf_\Xbf^{ij}  p_\smpn^-   + \Xbf^i \Xbf^j p_\smpn^-   - \half  \delta^{ij} \Xbf^2  p_\smpn^- + \delta^{ij} k_\smpn^- = 0 \,, \qquad\qquad
\\
\label{man-15122016-28} &&  \Kbf^i  k_\smpn^-   - \Xbf^i k_\smpn^- +  \Kbf_\Xbf^{il} j_\smpn^{-l}   - \Xbf^i \Xbf^l j_\smpn^{-l} + \half  \Xbf^2  j_\smpn^{-i}  = 0\,,
\eeq
where $\Xbf^i$ is defined in \rf{man-15122016-11} and we use the notation

\beq
\label{man-15122016-29} && \Kbf^i \equiv \sum_{a=1}^n \Big( \frac{1}{2} p_a^i(- \partial_{p_a^j} \partial_{p_a^j} +  M_a^\ommsm)
+  \big( \beta_a \partial_{\beta_a}  +  p_a^j \partial_{p_a^j} + \frac{d-2}{2} - M_a^\ompsm\big) \partial_{p_a^i}
\nonumber\\
&& \hspace{2cm}    - M_a^{ij}\partial_{p_a^j} + M_a^{\ominussm i}\Big)\,,
\\
&& \Kbf_\Xbf^{ij}  \equiv  \sum_{a=1}^n  \frac{1}{2n} M_a^\ommsm \delta^{ij}  + \frac{1}{n} (\partial_{p_a^i}\partial_{p_a^j} - \half \delta^{ij} \partial_{p_a^l} \partial_{p_a^l})\,.
\eeq

\noinbf{$J^{+i}$-kinematical symmetry restrictions}. Commutators between the dynamical generators \rf{man-14122016-07} and the kinematical generators $J^{+i}$ tell us that the densities $p_\smpn^-$, $j_\smpn^{-i}$, and $k_\smpn^-$ \rf{man-15122016-14}
depend on the momenta $p_a^i$ through the new momentum variables
$\Po_{ab}^i$ defined by
\be \label{man-15122016-32}
\Po_{ab}^i \equiv p_a^i \beta_b - p_b^i \beta_a\,.
\ee
In other words, the densities $p_\smpn^-$, $j_\smpn^{-i}$, $k_\smpn^-$ \rf{man-15122016-14}  turn out to be functions of $\Po_{ab}^i$ in place of $p_a^i$,
\be \label{man-15122016-33}
p_\smpn^- = p_\smpn^- (\Po_{ab},\beta_a;\, \alpha_a)\,, \qquad   j_\smpn^{-i} = j_\smpn^{-i} (\Po_{ab},\beta_a;\, \alpha_a)\,, \qquad   k_\smpn^- = k_\smpn^- (\Po_{ab},\beta_a;\, \alpha_a)\,.
\ee

We summarize our study of the dynamical generators \rf{man-14122016-07} by the following two remarks.

\noinbf{i})  The commutators between the dynamical generators \rf{man-14122016-07} and the kinematical
generators $J^{+-}$, $D$, $J^{ij}$, $K^+$, $K^i$  lead to equations \rf{man-15122016-17}-\rf{man-15122016-28}
for the densities $p_\smpn^-$, $j_\smpn^{-i}$, $k_\smpn^-$ \rf{man-15122016-14}.

\noinbf{ii}) The commutators between the dynamical generators \rf{man-14122016-07} and the kinematical
generators $J^{+i}$ tell us that the densities $p_\smpn^-$, $j_\smpn^{-i}$, $k_\smpn^-$ \rf{man-15122016-14} turn out to be functions of $\Po_{ab}^i$ in place of $p_a^i$ \rf{man-15122016-33}.

Using \rf{man-15122016-32}, we verify that by virtue of momentum conservation laws not all
momenta $\Po_{ab}^i$ are independent. Namely, we verify that only $n-2$ momenta $\Po_{ab}^i$ are independent.

\newsection{\large
Kinematical and dynamical equations for cubic vertices and light-cone gauge dynamical principle} \label{seccc-04}

First, in this section, we discuss the kinematical and dynamical equations for cubic vertices. Second, we discuss the light-cone gauge dynamical principle for cubic vertices.

The kinematical equations for $G_\smpn^\dyn$, $n\geq 3$, have already been discussed in the previous section.
We note then that, for $n=3$, there is only one independent $\Po_{ab}^i$ \rf{man-15122016-32}. This simplifies the kinematical equations for the dynamical generators $G_\smp3^\dyn$. Namely, using the momentum conservation laws,
\be \label{man-17122016-01}
p_1^i + p_2^i + p_3^i = 0\,, \qquad \quad \beta_1 +\beta_2 +\beta_3 =0 \,,
\ee
we verify that the momenta $\Po_{12}^i$, $\Po_{23}^i$, $\Po_{31}^i$ can be expressed in terms of a new momentum $\Po^i$ as
\be \label{man-17122016-02}
\Po_{12}^i =\Po_{23}^i = \Po_{31}^i = \Po^i \,,
\ee
where the new momentum $\Po^i$ is defined by the relations
\be \label{man-17122016-03}
\Po^i \equiv \frac{1}{3}\sum_{a=1,2,3}\check{\beta}_a p_a^i\,, \qquad
\check{\beta}_a\equiv \beta_{a+1}-\beta_{a+2}\,, \quad \beta_a\equiv
\beta_{a+3}\,.
\ee
We prefer to use the momentum $\Po^i$ \rf{man-17122016-03} because this momentum is manifestly
invariant under cyclic permutations of the external line indices
$1,2,3$. Therefore the densities $p_\smp3^-$, $j_\smp3^{-i}$, and $k_\smp3^-$ are eventually functions of
$\Po^i$, $\beta_a$ and $\alpha_a$:
\be \label{man-17122016-04}
p_\smp3^- = p_\smp3^-(\Po,\beta_a;\, \alpha_a)\,, \qquad
j_\smp3^{-i} = j_\smp3^{-i}(\Po,\beta_a;\, \alpha_a)\,, \qquad k_\smp3^- = k_\smp3^-(\Po,\beta_a;\, \alpha_a)\,.
\ee

\noinbf{Kinematical equations}. Before to present the kinematical equations we explain our notation. For an quantity $\chi$, we introduce the derivative $\partial_\chi$ and the operator $N_\chi$,
\be \label{man-17122016-17a}
\partial_\chi \equiv \partial /\partial \chi\,, \qquad N_\chi \equiv \chi \partial_\chi\,.
\ee
Various operators constructed out of the momenta $\Po^i$, $\beta_a$ and the $M$-operators are defined as
\beq
\label{man-17122016-17} &&   \Jbf^{ij} \equiv \Po^i \partial_{\Po^j}  -  \Po^j \partial_{\Po^i} + \Mbf^{ij}\,,
\\
\label{man-17122016-18} && \Dbf \equiv N_\Po - \Mbf^\ompsm + \frac{d-2}{2}\,,   \qquad N_\Po \equiv \Po^i\partial_{\Po^i}\,,
\\
\label{man-17122016-30} && \Kbf^+ \equiv \half\big(\beta\partial_{\Po^i}\partial_{\Po^i} +
\sum_{a=1,2,3} \beta_a  M_a^\ommsm \big)\,,
\\
\label{man-17122016-22} && \Kbf^i   =   (\No_\beta - \Mo^\ompsm)\partial_{\Po^i} - \Mo^{ij} \partial_{\Po^j} + \Mbf^{\ominussm i} - \frac{\Po^i}{6\beta}\sum_{a=1,2,3}
\beta_a \check\beta_a M_a^\ommsm\,,
\\
\label{man-17122016-23} && \Kbf_\Xbf^{ij} = \frac{1}{6} \delta^{ij} \Mbf^\ommsm
+ \frac{\Delta_\beta}{9} ( \partial_{\Po^i}\partial_{\Po^j}  - \half \delta^{ij} \partial_{\Po^l}\partial_{\Po^l})\,,
\\
\label{man-17122016-39} && \Pbf^-  =  \frac{\Po^i \Po^i}{2\beta} - \sum_{a=1,2,3} \frac{M_a^\oppsm}{2\beta_a}\,,
\\
\label{man-17122016-40} && \Jbf^{-i}  =    - \frac{\Po^i}{\beta} \No_\beta + \frac{1}{\beta} \Mo^{ij} \Po^j + \sum_{a=1,2,3}  \frac{\check\beta_a }{6\beta_a} M_a^\oppsm \partial_{\Po^i}  - \frac{1}{\beta_a} M_a^{\oplussm i}\,,
\\
\label{man-17122016-24} && \Mbf^{ij} \equiv \sum_{a=1,2,3} M_a^{ij}\,, \hspace{1.3cm} \Mo^{ij} \equiv \frac{1}{3} \sum_{a=1,2,3} \betach_a M_a^{ij}\,, \
\\
\label{man-17122016-24a} && \Mbf^\ompsm \equiv \sum_{a=1,2,3} M_a^\ompsm\,, \hspace{1cm} \Mo^\ompsm \equiv \frac{1}{3} \sum_{a=1,2,3} \check\beta_a M_a^\ompsm\,, \
\\
\label{man-17122016-25} && \Mbf^\ommsm \equiv \sum_{a=1,2,3} M_a^\ommsm\,, \hspace{1.1cm} \Mbf^{\omsm i} \equiv \sum_{a=1,2,3} M_a^{\omsm i}\,,
\\
\label{man-17122016-26} &&  \Nbf_\beta \equiv \sum_{a=1,2,3} \beta_a\partial_{\beta_a}\,, \hspace{1.4cm} \No_\beta   \equiv \frac{1}{3}\sum_{a=1,2,3} \check\beta_a \beta_a \partial_{\beta_a}\,,
\qquad
\\
\label{man-17122016-27} && \beta \equiv \beta_1\beta_2\beta_3\,, \qquad \betach \equiv \betach_1\betach_2\betach_3\,, \qquad \Delta_\beta \equiv \sum_{a=1,2,3} \beta_a^2\,. \qquad
\eeq
Using notation in \rf{man-17122016-17a}-\rf{man-17122016-27}, we now represent the kinematical equations given in \rf{man-15122016-17}-\rf{man-15122016-28} in terms of the densities \rf{man-17122016-04}.

\noinbf{$J^{+-}$-kinematical equations}:

\vspace{-5pt}
\be \label{man-17122016-05}
\big(N_\Po + \Nbf_\beta \big)  p_\smp3^-  = 0\,, \qquad \big( N_\Po + \Nbf_\beta \big)\, j_\smp3^{-i}  = 0\,, \qquad  \big( N_\Po + \Nbf_\beta \big)\, k_\smp3^-  = 0\,;\qquad
\ee

\noinbf{$D$-kinematical equations}:

\vspace{-5pt}
\be \label{man-17122016-08}
(\Dbf -2) p_\smp3^-  = 0\,, \qquad (\Dbf - 1 )  j_\smp3^{-i} = 0 \,, \qquad   \Dbf  k_\smp3^- = 0\,;
\ee

\noinbf{$J^{ij}$-kinematical equations}:

\vspace{-5pt}
\be \label{man-17122016-14}
\Jbf^{ij}  p_\smp3^- =0\,, \qquad\Jbf^{ij} j_\smp3^{-k} = \delta^{jk} j_\smp3^{-i} - \delta^{ik} j_\smp3^{-j}\,, \hspace{1cm}  \Jbf^{ij} k_\smp3^- =0\,;
\ee

\noinbf{$K^+$-kinematical equations}:

\vspace{-10pt}
\be \label{man-17122016-28}
\Kbf^+   p_\smp3^- =  0 \,, \qquad  \Kbf^+   j_\smp3^{-i} =  0 \,,\hspace{1cm} \Kbf^+   k_\smp3^- =  0 \,;
\ee

\noinbf{$K^i$-kinematical equations}:

\vspace{-25pt}
\beq
\label{man-17122016-19} &&    j_\smp3^{-i} -  \Kbf^i  p_\smp3^-   = 0 \,,
\\
\label{man-17122016-20} &&  \Kbf^i  j_\smp3^{-j} -  \Kbf_\Xbf^{ij} p_\smp3^-     + \delta^{ij}  k_\smp3^- = 0 \,, \qquad
\\
\label{man-17122016-21} &&  \Kbf^i  k_\smp3^-     +  \Kbf_\Xbf^{il} j_\smp3^{-l}     = 0\,.
\eeq

Note that, for the derivation of equations \rf{man-17122016-08}, we use equations  \rf{man-17122016-05}.
For the derivation of equations \rf{man-17122016-19}-\rf{man-17122016-21}, we use equations \rf{man-15122016-26}-\rf{man-15122016-28} and the relation $[\Xbf^i,\Po^j]=0$.
Equations given in \rf{man-17122016-05}-\rf{man-17122016-21} constitute the full list of kinematical equations.

\noinbf{Dynamical equations}. The dynamical equations for vertices are obtained by using commutators between dynamical generators \rf{man-14122016-07}. The full list of those commutators is given by
\beq
\label{man-17122016-31} && [P^-,J^{-i}] = 0 \,,
\\
\label{man-17122016-32} && [P^-,K^-] = 0 \,, \qquad [J^{-i},J^{-j}] = 0 \,, \qquad [J^{-i}, K^-] = 0\,.
\eeq
We note however that we can restrict ourselves to the study of the commutator in \rf{man-17122016-31} in view of the following important feature of the conformal algebra. Kinematical $K^i$-symmetry equations given in \rf{man-17122016-19}-\rf{man-17122016-21} are obtained from the commutators
\be \label{man-17122016-33}
[P^-,K^i] = -J^{-i}\,, \qquad [J^{-i},K^j] = \delta^{ij} K^- \,, \qquad [K^-,K^i] = 0\,.
\ee
Using Jacoby identities, we verify that, if the commutators \rf{man-17122016-31} and \rf{man-17122016-33} are satisfied, then the commutators \rf{man-17122016-32} are satisfied automatically. Thus, if we respect $K^i$-symmetry equations \rf{man-17122016-19}-\rf{man-17122016-21}, then we can indeed restrict ourselves to the study of the commutator in \rf{man-17122016-31}.

To cubic order in fields, the commutator \rf{man-17122016-31} takes the form
\be \label{man-17122016-34}
[P_\smpt^- \,,J_\smp3^{-i}] + [P_\smp3^-\,,J_\smpt^{-i}]=0\,.
\ee
Equations \rf{man-17122016-34} amount to the following equations for
$p_\smp3^-$ and $j_\smp3^{-i}$ \rf{man-17122016-04},
\be \label{man-17122016-35}
\Jbf^{-i}  p_\smp3^-  + \Pbf^-   j_\smp3^{-i}  = 0\,,
\ee
where $\Pbf^-$ and $\Jbf^{-i}$ are given in \rf{man-17122016-39}, \rf{man-17122016-40}. Dynamical equations \rf{man-17122016-35} involve both the cubic vertex $p_\smp3^-$ and the density $j_\smp3^{-i}$.
Note however that the kinematical equation \rf{man-17122016-19} allows us to express $j_\smp3^{-i}$ in terms of $p_\smp3^-$. Plugging $j_\smp3^{-i}$ \rf{man-17122016-19} into \rf{man-17122016-35}, we obtain the {\it closed} equations for the cubic vertex $p_\smp3^-$,
\be \label{man-17122016-41}
\bigl( \Jbf^{-i}  + \Pbf^- \Kbf^i\bigr)  p_\smp3^-  = 0\,.
\ee
The possibility to derive the closed equations \rf{man-17122016-41} for the cubic vertex by using only commutators of the conformal algebra is the attractive feature of the light-cone approach to conformal fields.%
\footnote{For the case of Poincar\'e algebra symmetries, closed equations for cubic vertex can be obtained only after a choice of a representative for cubic vertex. An example of the closed equations for so called harmonic representative of cubic vertex is given in Sec. 4.1 in Ref.\cite{Metsaev:2005ar}.}

Thus, to cubic order in fields, we exhaust all commutators of the conformal algebra. Dynamical equations \rf{man-17122016-41} and kinematical equations  \rf{man-17122016-05}-\rf{man-17122016-21} provide the complete list of restrictions imposed by
commutators of the conformal algebra on the densities $p_\smp3^-$,
$j_\smp3^{-i}$, and $k_\smp3^-$.

\noindent {\bf Light-cone gauge dynamical principle}. Kinematical equations \rf{man-17122016-05}-\rf{man-17122016-21} and dynamical equation \rf{man-17122016-41} by themselves do not provide a possibility to fix all solutions for cubic vertex uniquely. To determine all solutions for cubic vertex uniquely we impose so called light-cone dynamical principle.
We formulate this principle as the following two requirements:
\beq
\label{man-17122016-42} && \hspace{-8cm} \abf)\qquad  p_\smp3^-\,, \  j_\smp3^{-i}\,, \ k_\smp3^- \hspace{0.4cm} \hbox{ are polynomial in $\Po^i$};
\\
\label{man-17122016-43} && \hspace{-8cm}  \bbf)\qquad p_\smp3^-  \ne  \Pbf^- W\,, \quad W \ \hbox{is polynomial in } \Po^i;
\eeq
where $\Pbf^-$ is given in \rf{man-17122016-39}. We recall that, upon field redefinitions, the cubic vertex $p_\smp3^-$ is changed as $p_\smp3^- \rightarrow p_\smp3^- + \Pbf^- f$ (see, e.g., Appendix B in Ref.\cite{Metsaev:2005ar}). Therefore, if we ignore the restriction \rf{man-17122016-43}, then, among other solutions, we can meet the solutions for $p_\smp3^-$  which can be removed by the field redefinitions. As we prefer to deal with the cubic vertex $p_\smp3^-$
that cannot be removed by the field redefinitions we impose the restriction \rf{man-17122016-43}. Note also that, if some solution for the vertex $p_\smp3^-$ obeys our kinematical and dynamical equations, then the vertex $p_\smp3^-{}^f$ obtainable from $p_\smp3^-$ by using the field redefinitions, $p_{\smp3}^-{}^f = p_\smp3^-  + \Pbf^- f$, also obeys our equations. Therefore to find all solutions for the cubic vertex uniquely we choose some representative of the vertex by exploiting the field redefinitions.  After choosing a representative of the cubic vertex our kinematical and dynamical equations allow us to find all solutions for the cubic vertex $p_\smp3^-$ and the corresponding densities $j_\smp3^{-i}$ and $k_\smp3^-$ uniquely. Below, we use our general setup above described for the detailed study of cubic vertices for scalar and vector fields.

\newsection{ \large
Scalar and vector fields. Equations for representative of cubic vertices} \label{seccc-05}

The cubic vertex  \rf{man-17122016-04} can be written as
\be \label{man-16082022-01}
p_\smp3^- = p_\smp3^-(\Po^i\,, \alpha_a^i\,, \zeta_a\,, \upsilon_a^\opsm\,,\upsilon_a^\omsm\,,\beta_a)\,,
\ee
where we show all arguments explicitly.
In view of the $so(d-2)$ symmetries \rf{man-17122016-14}, a general parity-even and $so(d-2)$ invariant solution for the cubic vertex \rf{man-16082022-01} can be presented as%
\footnote{ In \rf{man-16082022-02}, we ignore the $so(d-2)$ algebra invariants that can be built by using the Levi-Civita antisymmetric tensor $\epsilon^{i_1\ldots i_{d-2}}$. We refer to vertices  not involving the Levi-Civita antisymmetric tensor as parity-even vertices.}
\be \label{man-16082022-02}
p_\smp3^- =  p_\smp3^-(L_a, \alpha_{aa+1}, \zeta_a\,, \upsilon_a^\omsm\,, \upsilon_a^\opsm\,, \beta_a\,, \Po^2)\,,
\ee
where, in \rf{man-16082022-02} and below, we use the following notation for various variables:
\be \label{man-16082022-03}
L_a \equiv \frac{\Po^i\alpha_a^i}{\beta_a}\,, \qquad \alpha_{ab}\equiv \alpha_a^i\alpha_b^i\,, \qquad \Po^2 \equiv \Po^i\Po^i\,,\qquad \upsilonbf^\opsm \equiv \sum_{a=1,2,3} \upsilon_a^\opsm\,.\qquad
\ee
In other words, the vector oscillators $\alpha_a^i$ and the momentum $\Po^i$ enter the cubic vertex \rf{man-16082022-02} through the $so(d-2)$ symmetry invariants $\alpha_a^i\Po^i$, $\alpha_a^i \alpha_{a+1}^i$, and $\Po^i\Po^i$. Note however that we find it convenient,  in place of the invariant $\alpha_a^i\Po^i$,  to use the invariant $L_a$ defined in \rf{man-16082022-03}.

Up to now all our equations were valid for arbitrary spin conformal fields. From now on we restrict our attention to scalar and vector conformal fields.
In Appendix A, we prove the following

\noinbf{ Statement}. For scalar and vector conformal fields, by using field redefinitions, the vertex $p_\smp3^-$ \rf{man-16082022-02} can be made independent of $\Po^2$,
\be \label{man-16082022-04}
p_\smp3^- =  p_\smp3^-(L_a, \alpha_{aa+1}, \zeta_a\,, \upsilon_a^\omsm\,, \upsilon_a^\opsm\,, \beta_a)\,.
\ee
In other words, for scalar and vector conformal fields, without loss of generality, the representative of the cubic vertex $p_\smp3^-$ can be chosen as in \rf{man-16082022-04}.

For the representative of the cubic vertex \rf{man-16082022-04}, our equations \rf{man-17122016-05}-\rf{man-17122016-28} and
\rf{man-17122016-41} allow us to find all possible solutions for the cubic vertices uniquely. The representative of the cubic vertex depends on the variables shown explicitly in \rf{man-16082022-04}. Therefore it is reasonable to represent various differential operators entering our equations in terms of the variables appearing in \rf{man-16082022-04}.

\noinbf{Realization of the operators $\Dbf$, $\Kbf^+$, $\Kbf^i$, and $\Jbf^{-i}$ on the representative of $p_\smp3^-$ \rf{man-16082022-04} for scalar and vector fields}. On space of $p_\smp3^-$ \rf{man-16082022-04}, the operators $\Dbf$, $\Kbf^+$, $\Kbf^i$, and $\Jbf^{-i}$ given in \rf{man-17122016-18}-\rf{man-17122016-40} are realized as follows,
\beq
\label{man-16082022-05a} && \Dbf = \Nbf_L - \Mbf^\ompsm + \frac{d-2}{2}\,,
\\
\label{man-16082022-05} &&   \Kbf^+ = \half \sum_{a=1,2,3} \beta_a \big( M_a^\ommsm  + 2 \alpha_{a+1a+2}\partial_{L_{a+1}} \partial_{L_{a+2}} \big) \,,
\\
\label{man-16082022-06}  &&    \Kbf^i      =    \Po^i G_\beta^\Kbf + \sum_{a=1,2,3} \alpha_a^i G_a^\Kbf\,,
\\
\label{man-16082022-07}  &&   \Jbf^{-i}     =   \Po^i G_\beta^\Jbf + \sum_{a=1,2,3} \frac{\alpha_a^i}{\beta_a} G_a^\Jbf +  \Pbf^- \sum_{a=1,2,3} \frac{2\betach_a}{3\beta_a}\alpha_a^i \partial_{L_a}\,,
\\
\label{man-16082022-08} && \Jbf^{-i} + \Pbf^- \Kbf^i =   \Po^i G_\beta^\Jbf + \sum_{a=1,2,3} \frac{\alpha_a^i}{\beta_a} G_a^\Jbf +   \Pbf^- \Big( \Po^i G_\beta^\Kbf + \sum_{a=1,2,3} \alpha_a^i G_a^{\Jbf\Kbf} \Big) \,, \qquad
\eeq
where operators $G_{a,\beta}^\Jbf$, $G_{a,\beta}^\Kbf$, and $G_a^{\Jbf\Kbf}$ are define as
\beq
\label{man-16082022-21} && G_\beta^\Kbf  =   \sum_{a=1,2,3}\frac{1}{\beta_a} m_a^\omsm \partial_{L_a} - \frac{\beta_a \betach_a}{3\beta}   \alpha_{a+1a+2} \partial_{L_{a+1}} \partial_{L_{a+2}}
- \frac{\beta_a \betach_a}{6\beta}   M_a^\ommsm\,,\qquad
\\
\label{man-16082022-22} && G_a^\Kbf   =  \frac{1}{\beta_a} \Big( \No_\beta - \No_L - \Mo^\ompsm  - \frac{1}{3}\betach_a ( \Nbf_L + d - 2) \Big) \partial_{L_a}
\nonumber\\
&& \hspace{0.8cm} +\,\,  \mb_a^\omsm + m_{a+1}^\omsm \partial_{\alpha_{aa+1}} +  m_{a+2}^\omsm \partial_{\alpha_{a+2a}} + \alpha_{a+1 a+2}  \big( \partial_{\alpha_{aa+1}} \partial_{L_{a+2}} -  \partial_{\alpha_{a+2a}} \partial_{L_{a+1}} \big)\,,\qquad
\\
\label{man-16082022-23} && G_\beta^\Jbf  =  - \frac{1}{\beta}  \No_\beta - \sum_{a=1,2,3}\frac{1}{\beta_a^2} m_a^{\opsm}\partial_{L_a}\,,
\\
\label{man-16082022-24} && G_a^\Jbf   =    \big( L_{a+2} - \frac{\beta_a}{\beta_{a+2}} m_{a+2}^\opsm \big) \partial_{\alpha_{a+2 a}} - \big( L_{a+1} + \frac{\beta_a}{\beta_{a+1}} m_{a+1}^\opsm \big) \partial_{\alpha_{aa+1}}
\nonumber\\
&& \hspace{0.8cm} +\,\,   \half \Bigl( \frac{\betach_a}{ \beta_a} M_a^\oppsm +  M_{a+1}^\oppsm - M_{a+2}^\oppsm  \Bigr) \partial_{L_a} - \mb_a^\opsm\,,
\\
\label{man-16082022-25}  &&  G_a^{\Jbf\Kbf}   =  \frac{1}{\beta_a} \Big( \No_\beta - \No_L - \Mo^\ompsm  - \frac{1}{3}\betach_a ( \Nbf_L + d - 4) \Big) \partial_{L_a}
\nonumber\\
&& \hspace{1cm} + \,\, \mb_a^\omsm + m_{a+1}^\omsm \partial_{\alpha_{aa+1}} +  m_{a+2}^\omsm \partial_{\alpha_{a+2a}}
+ \alpha_{a+1 a+2}  \big( \partial_{\alpha_{aa+1}} \partial_{L_{a+2}} -  \partial_{\alpha_{a+2a}} \partial_{L_{a+1}} \big)\,,\qquad\qquad
\\
\label{man-17122016-26a} && \hspace{1cm} \Nbf_L = \sum_{a=1,2,3} N_{L_a}\,,\qquad \No_L = \frac{1}{3}\sum_{a=1,2,3} \betach_a N_{L_a}\,, \qquad N_{L_a}\equiv L_a \partial_{L_a}\,.
\eeq
For the notation see also \rf{man-17122016-24a}-\rf{man-17122016-27}. For scalar and vector fields, the operators $m^\opsm$, $m^\omsm$ and $\mb^\opsm$, $\mb^\omsm$  are given in \rf{man-14122016-67}, while the operators $M^\ompsm$, $M^\oppsm$, $M^\ommsm$ are given in \rf{man-14122016-42}.  Using the notation in \rf{man-17122016-26}, \rf{man-16082022-05a}, \rf{man-16082022-05}, and \rf{man-16082022-21}-\rf{man-17122016-26a}, we now summarize our

\noinbf{Complete system of equations for representative of cubic vertex \rf{man-16082022-04}}:
\beq
\label{man-16082022-31} && \Nbf_\beta p_\smp3^- = 0\,,
\\
\label{man-16082022-32} && \bigl( \Dbf   -  2 \bigr) p_\smp3^-  = 0 \,,
\\
\label{man-16082022-33} && \Kbf^+ \, p_\smp3^-= 0 \,,
\\
\label{man-16082022-34} && G_\beta^\Jbf \, p_\smp3^-  = 0 \,,
\\
\label{man-16082022-35} && G_\beta^\Kbf \, p_\smp3^- = 0 \,,
\\
\label{man-16082022-36} && G_a^{\Jbf\Kbf} \, p_\smp3^- = 0 \,, \hspace{1cm} a=1,2,3;
\\
\label{man-16082022-37} && G_a^\Jbf \, p_\smp3^-  = 0 \,, \hspace{1.2cm} a=1,2,3\,.
\eeq
Equations \rf{man-16082022-31}, \rf{man-16082022-32}, and \rf{man-16082022-33} are obtained by plugging the representative of $p_\smp3^-$ \rf{man-16082022-04} into the respective equations \rf{man-17122016-05}, \rf{man-17122016-08} and \rf{man-17122016-28}, while equations \rf{man-16082022-34}-\rf{man-16082022-37} are obtained from equations \rf{man-17122016-41} by using the relation in \rf{man-16082022-08}. For the reader convenience, we note also the following two sets of algebraic constraints for the representative of $p_\smp3^-$ \rf{man-16082022-04}:
\beq
\label{man-16082022-38} && \big( N_{L_a} + N_{\alpha_{aa+1}} + N_{\alpha_{a+2a}} + N_{\zeta_a} - s_a \big)p_\smp3^- = 0 \,, \hspace{1.5cm} a=1,2,3;
\\
\label{man-16082022-39} && \big( N_{\upsilon_a^\opsm} + N_{\upsilon_a^\omsm} + N_{\zeta_a} - k_{s_a} \big)p_\smp3^- = 0 \,,   \hspace{3.4cm} a=1,2,3\,.\qquad
\eeq
Equations \rf{man-16082022-38}, \rf{man-16082022-39} tell us that the cubic vertex $p_\smp3^-$ describes an interaction of fields which satisfy the respective first and second algebraic constraints in \rf{man-14122016-29}.

Equations \rf{man-16082022-31}-\rf{man-16082022-39} allow us to find all solutions for the cubic vertex uniquely. These solutions are discussed in the next section, while the systematic method for solving equations \rf{man-16082022-31}-\rf{man-16082022-39} is presented in Appendix B. Note however that, besides the cubic vertex, we have to fix the densities $j_\smp3^{-i}$ and $k_\smp3^-$. These densities can be fixed by using the $K^i$-symmetry equations given in \rf{man-17122016-19}-\rf{man-17122016-21}. Here we present our result (for details of the derivation, see Appendix C),
\beq
\label{man-16082022-40} && j_\smp3^{-i} =  - \sum_{a=1,2,3}\frac{2\betach_a}{3\beta_a} \alpha_a^i\partial_{L_a}^{\vphantom{8pt}} p_\smp3^-\,,
\\
\label{man-16082022-41} && k_\smp3^- =   \sum_{a=1,2,3} \Big( \frac{2\betach_a}{3\beta_a} m_a^\omsm \partial_{L_a}^{\vphantom{8pt}} + \frac{2}{9} \big(\frac{4\beta_a^3}{\beta}-1\big)
\alpha_{a+1a+2}  \partial_{L_{a+1}}^{\vphantom{8pt}}  \partial_{L_{a+2}}^{\vphantom{8pt}}\Big) p_\smp3^-   +  \frac{1}{6} \Mbf^\ommsm p_\smp3^-\,,\qquad\qquad
\eeq
where $m_a^\omsm$, $\Mbf^\ommsm$ are defined in \rf{man-14122016-67}, \rf{man-17122016-25}. From \rf{man-16082022-40}, \rf{man-16082022-41},  we see that the knowledge of the cubic vertex $p_\smp3^-$ provides us automatically the solution for the densities $j_\smp3^{-i}$ and $k_\smp3^-$. To summarize, all that remains is to find the cubic vertex $p_\smp3^-$.

\newsection{ \large Cubic vertices for scalar and vector fields }\label{seccc-06}

We present our solution for cubic vertex $p_\smp3^-$ \rf{man-16082022-04} in terms of a dressing operator denoted as $U$ and a undressed vertex denoted as $V$ (for details of the derivation, see Appendix B),
\be \label{man-21082022-01}
p_\smp3^- = UV\,, \hspace{1cm}  V = V(L_a,\alpha_{aa+1},\upsilonbf^\opsm)\,,
\ee
where we show that the undressed vertex $V$ depends only on the variables $L_a$, $\alpha_{aa+1}$, and $\upsilonbf^\opsm$ defined in \rf{man-16082022-03}. The undressed vertex $V$ satisfies the following equations:
\beq
\label{man-21082022-03} && \hspace{-1cm} G_a V  = 0 \,, \qquad G_a \equiv L_{a+2}  \partial_{\alpha_{a+2 a}}  - L_{a+1}  \partial_{\alpha_{aa+1}}  +      \upsilonbf^\opsm  u_{\zeta,a}  \,,\qquad a=1,2,3\,,
\\
\label{man-21082022-04} && \hspace{-1cm} \bigl( \Nbf_L    + 2N_{\upsilonbf^\opsm}  - \kbf_s + \frac{d-6}{2} \bigr) V  = 0 \,,
\\
\label{man-21082022-05} && \hspace{-1cm} \big( N_{L_a} + N_{\alpha_{aa+1}} + N_{\alpha_{a+2a}} - s_a \big) V  = 0 \,, \hspace{4.2cm} a=1,2,3;
\\
\label{man-21082022-06} && \hspace{1cm} \kbf_s \equiv \sum_{a=1,2,3} k_{s_a}\,,
\nonumber\\
\label{man-21082022-07} && \hspace{1cm}  k_{s_a} =  k_a\,, \quad k_a \in \No_0\,,\hspace{0.6cm}  \for \quad s_a =0\,,
\nonumber\\
\label{man-21082022-08} && \hspace{1cm} k_{s_a} =  \frac{d-4}{2}\,, \hspace{2cm} \for \quad s_a =1\,;
\eeq
and the following two requirements:
\beq
\label{man-21082022-09} && \hspace{-4cm} \abf) \hspace{0.5cm} V \ \hbox{ is polynomial in } \ L_a\,, \ \alpha_{aa+1}\,, \ \upsilonbf^\opsm\,, \quad a=1,2,3\,;
\\
\label{man-21082022-09a} &&  \hspace{-4cm}  \bbf) \hspace{0.5cm}  V^\omsm \ \hbox{ is polynomial in } \  \upsilon_1^\omsm\,, \upsilon_2^\omsm\,, \upsilon_3^\omsm\,,\hspace{1cm} V^\omsm \equiv  \prod_{a=1,2,3} (\upsilon_a^\omsm)^{k_{s_a} - N_{\upsilon_a^\opsm}} V\,;
\eeq
where $u_{\zeta,a}$ \rf{man-21082022-03} is defined below in \rf{man-21082022-15}. The dressing operator $U$ can be presented as
\beq
\label{man-21082022-10} && U  =    U_\beta U_{\upsilon^\omsm} U_{\upsilon^\opsm} U_{N_{\upsilon^\opsm}} U_\zeta U_{\zeta\zeta} \,,
\\
\label{man-21082022-11} && U_\beta = e^{u_\beta}\,, \qquad u_\beta =  -\sum_{a=1,2,3}\frac{\betach_a}{2\beta_a} m_a^\opsm \partial_{L_a}\,,
\\
\label{man-21082022-12} && U_{\upsilon^\omsm} \equiv \prod_{a=1,2,3} U_{\upsilon_a^\omsm}\,, \hspace{1cm} U_{\upsilon_a^\omsm} \equiv (\upsilon_a^\omsm)^{ k_{s_a} - N_{\upsilon_a^\opsm} - N_{\zeta_a} }\,,
\\
\label{man-21082022-13} && U_{\upsilon^\opsm} = e^{u_{\upsilon^\opsm}}\,, \hspace{1cm} u_{\upsilon^\opsm} = \sum_{a=1,2,3} u_{\upsilon\, a} \upsilon_a^\opsm\,,
\nonumber\\
&& \hspace{2cm} u_{\upsilon\, a} = - \half \alpha_{a+1a+2}\partial_{L_{a+1}}\partial_{L_{a+2}}
+ \frac{1}{4} \big(  \alpha_{aa+1}\zeta_{a+2}  -  \alpha_{a+2a}\zeta_{a+1} \big) \partial_{L_1}\partial_{L_2}\partial_{L_3}\,,
\\
\label{man-21082022-14} && U_{N_{\upsilon^\opsm}} = \prod_{a=1,2,3} U_{N_{\upsilon_a^\opsm}}\,, \hspace{1cm} U_{N_{\upsilon_a^\opsm}} = r_a{}^{\!\! N_{\upsilon_a^\opsm}}\,,
\nonumber\\
&& \hspace{2cm}   r_a =  1 + \half   \zeta_{a+1}\partial_{L_{a+1}} - \half \zeta_{a+2} \partial_{L_{a+2}}  + \frac{3}{4} \zeta_{a+1} \zeta_{a+2} \partial_{L_{a+1}} \partial_{L_{a+2}}\,,
\\
\label{man-21082022-15} && U_\zeta = e^{u_\zeta}\,, \hspace{1cm} u_\zeta  =  \sum_{a=1,2,3} u_{\zeta,a}\zeta_a \,,
\nonumber\\
&& \hspace{2cm} u_{\zeta,a} \equiv \half \big( N_{L_{a+1}} - N_{L_{a+2}} + \half \kb_{s_{a+1}} - \half \kb_{s_{a+2}}\big)\partial_{L_a}\,,\qquad \kb_{s_a} \equiv k_{s_a} - s_a\,,\qquad\qquad
\\
\label{man-21082022-16} && U_{\zeta\zeta} = e^{u_{\zeta\zeta}}\,,\qquad u_{\zeta\zeta}=\sum_{a=1,2,3}u_{\zeta\zeta,a} \zeta_{a+1}\zeta_{a+2}\,,
\nonumber\\
&& \hspace{3cm} u_{\zeta\zeta,a}  =   \frac{1}{8}\big( \frac{d-2}{2} - \kb_{s_a}   \big)  \partial_{L_{a+1}} \partial_{L_{a+2}}
 - \partial_{ \alpha_{a+1a+2} } \partial_{\upsilonbf^\opsm} \,,\qquad
\eeq
where the operators $m^\opsm$, $m^\omsm$ and $\mb^\opsm$, $\mb^\omsm$  are given  in \rf{man-14122016-67}. Operators $u_{\zeta\zeta,a}$ \rf{man-21082022-16} can alternatively be represented as (see \rf{09072022-04}, \rf{09072022-06} in Appendix B),
\be \label{man-21082022-16a}
u_{\zeta\zeta,a}  =   \frac{1}{4}\big(  \frac{d-4}{2} - N_{\upsilonbf^\opsm} \big)  \partial_{L_{a+1}} \partial_{L_{a+2}}
- \partial_{ \alpha_{a+1a+2} } \partial_{\upsilonbf^\opsm}\,.
\ee

By definition, the vertex $p_\smp3^-$ should be polynomial in all oscillators.  Using the dressing operator $U$ \rf{man-21082022-10},  we see that, if the undressed vertex $V$ satisfies requirement \rf{man-21082022-09}, then the vertex $p_\smp3^-$, with the exception of the oscillators $v_1^\omsm$, $v_2^\omsm$, $v_3^\omsm$, is obviously polynomial in all oscillators. In order to get vertex $p_\smp3^-$ that is polynomial in the oscillators $v_1^\omsm$, $v_2^\omsm$, $v_3^\omsm$, we impose restriction \rf{man-21082022-09a}. This restriction can be represented in easy-to-use form by using the Taylor series expansion of the undressed vertex $V$ in $\upsilonbf^\opsm$,
\be \label{man-21082022-17}
V = \sum_{m = m_\minrm}^{m_\maxrm} V_m \upsilonbf^{\opsm\, m}\,, \qquad V_m = V_m(L_a,\alpha_{aa+1})\,,
\ee
where non-negative integers $m_\minrm$ and $m_\maxrm$ depend on spins and conformal dimensions and are fixed by using equations \rf{man-21082022-03}-\rf{man-21082022-05}.
Plugging $V$ \rf{man-21082022-17} into \rf{man-21082022-09a}, we get $V^\omsm$ given by:
\be \label{man-21082022-18}
V^\omsm  = \sum_{m=m_\minrm}^{m_\maxrm} \upsilon^m V_m \prod_{a=1,2,3} \upsilon_a^{\omsm\,\, k_{s_a}-m}\,, \qquad  \upsilon \equiv \upsilon_1^\oplussm \upsilon_2^\ominussm \upsilon_3^\ominussm
+ \upsilon_2^\oplussm \upsilon_3^\ominussm \upsilon_1^\ominussm
+ \upsilon_3^\oplussm \upsilon_1^\ominussm \upsilon_2^\ominussm\,.\qquad
\ee
From \rf{man-21082022-18}, we see that requirement \rf{man-21082022-09a} amounts to the restrictions
\be
k_{s_a}-m \geq 0\,, \qquad 0 \leq m_\minrm \leq m \leq m_\maxrm\,, \qquad a=1,2,3\,.
\ee

Finding all solutions to equations \rf{man-21082022-03}-\rf{man-21082022-05} and requirements \rf{man-21082022-09}, \rf{man-21082022-09a}  turns out to be simple problem. We now present our solutions for the undressed vertices.

\noindent {\bf Undressed vertex for three scalar fields}. Using notation as in \rf{man-20092022-01}, we consider cubic vertex for three scalar fields,
\beq
\label{man-21082022-20} && (\Delta_1,0) - (\Delta_2,0) - (\Delta_3,0)
\nonumber\\
&& \Delta_a = \frac{d-2}{2} - k_a \,, \qquad k_a \in \No_0\,, \qquad a=1,2,3\,,
\eeq
i.e. three scalar fields carry external line indices $a=1,2,3$.
For this case, we find one solution for the undressed vertex given by
\be \label{man-21082022-21}
V = \upsilonbf^{\opsm\,\, n}\,, \qquad n \equiv \half(\frac{6-d}{2} + \kbf) \,, \qquad \kbf\equiv \sum_{a=1,2,3} k_a\,,
\ee
where $n$ \rf{man-21082022-21} and $n_a$ defined in \rf{man-21082022-22a} should satisfy the restrictions given by
\beq
&& \label{man-21082022-22}   n \geq 0 \,, \ \qquad n \in \No_0 \,,
\\
&& \label{man-21082022-22a} n_a \geq 0\,, \qquad n_a \in \No_0 \,, \hspace{1cm} n_a \equiv \half \Bigl( \frac{d-6}{2} + k_a - k_{a+1} - k_{a+2} \Bigr)\,,\qquad a=1,2,3\,.\qquad \qquad
\eeq

The following remarks are in order.

\noinbf{i}) Restrictions \rf{man-21082022-22} are obtainable by requiring the power of $\upsilonbf^\opsm$ in \rf{man-21082022-21} be non--negative integer, while restrictions \rf{man-21082022-22a} are obtainable by using the requirement in \rf{man-21082022-09a}.

\noinbf{ii})  Restrictions \rf{man-21082022-22}, \rf{man-21082022-22a} amount to the following restrictions:
\be \label{man-21082022-23}
\kbf - 2k_\minrm\, \leq\, \half (d-6) \,  \leq \,  \kbf\,, \qquad \kbf \equiv \sum_{a=1,2,3} k_a\,, \qquad k_\minrm = \min_{a=1,2,3} k_a\,.
\ee
Restrictions \rf{man-21082022-23} lead to a surprisingly
simple result for allowed values of space-time dimensions.
Indeed, taking into account that $n_a\in \No_0$, we obtain that given
values $k_1$, $k_2$, $k_3$, the allowed values of  space-time dimension $d$ are given by
\be \label{man-21082022-24}
d = 2 \kbf+6,\ 2\kbf+2,\ 2\kbf-2, \ \ldots, 2\kbf+6 - 4k_\minrm\,.
\ee
Relation \rf{man-21082022-24} implies that given values $k_1$, $k_2$,
$k_3$, the number of allowed values of space-time dimensions $d$ which admit conformal invariant cubic vertices for scalar fields having conformal dimensions as in \rf{man-21082022-20} is given by
\be  \label{man-21082022-25}
k_\minrm + 1\,.
\ee
Besides this, relation \rf{man-21082022-24} implies that, for $d=4$, conformal invariant cubic interaction for the scalar fields does not exist.
We note also the interesting similarity between relations \rf{man-21082022-23}, \rf{man-21082022-24}, and \rf{man-21082022-25} in this paper and the respective relations (5.13), (5.15), and (5.16) in Ref.\cite{Metsaev:2005ar}.
Note also that the restrictions \rf{man-21082022-22a} can alternatively be represented as
\be \label{man-21082022-29}
k_1 + k_2 - \half (d-6) \leq k_3 \leq \half (d-6) - |k_1-k_2|\,.
\ee

\noinbf{iii}) Free action \rf{man-15122016-04} for the conformal {\it scalar} fields exists for both even $d$ and odd $d$.  From \rf{man-21082022-24}, we see however that cubic vertex for the scalar fields is available only for even $d$.

\noinbf{iv}) To illustrate \rf{man-21082022-24}, \rf{man-21082022-25} consider three scalar fields having canonical dimensions, $k_a=0$, $a=1,2,3$. From \rf{man-21082022-25} we see then that there is only one allowed value of $d$, while, from \rf{man-21082022-24}, we learn that the allowed value of $d$ is equal to 6. Note that, for the scalar fields in $R^{5,1}$, we get the dressing operator $U=1$ and therefore the undressed vertex $V$ coincides with the cubic vertex, $p_\smp3^-=V$.

\noinbf{v}) From \rf{man-21082022-24}, we get the restriction
\be  \label{man-21082022-26}
d\leq 2 \kbf + 6\,.
\ee
Restriction \rf{man-21082022-26} can easily be explained by using higher-derivative approach. In such approach, the scalar field having conformal dimension $\Delta = \frac{d-2}{2}-k$ is described by the field $\phi_{-k}$ entering \rf{man-14122016-08} with $k'=-k$. A higher-derivative Lorentz covariant cubic Lagrangian for three scalar fields $\phi_{-k_a}$ with conformal dimensions as in \rf{man-21082022-20} can schematically be presented as
\be   \label{man-21082022-27}
\LL_\smp3^{\rm high-deriv} = \partial^{l_1} \phi_{-k_1} \partial^{l_2} \phi_{-k_2} \partial^{l_3} \phi_{-k_3}\,,
\ee
where $\partial^l$ stands for $l$ derivatives. Requiring the dilatation symmetry of the Lagrangian \rf{man-21082022-27} gives the relation
\be  \label{man-21082022-28}
d =  2 \kbf + 6 - \sum_{a=1,2,3} l_a\,.
\ee
Taking into account that $l_a\geq 0$, we see that relation \rf{man-21082022-28} implies restriction \rf{man-21082022-26}. To our knowledge, though the higher-derivative Lorentz covariant cubic Lagrangian for the scalar fields is well known, the restrictions \rf{man-21082022-23} have not been reported in the earlier literature.

\noinbf{vi}) Undressed vertex V \rf{man-21082022-21} and the corresponding cubic vertex $p_\smp3^-$ \rf{man-21082022-01} do not depend on the momenta $\Po^i$. In general, cubic vertex for three scalar fields could depend on $\Po^i\Po^i$. However, as we have shown, using field redefinitions, the dependence of the cubic vertex on $\Po^i\Po^i$ can be removed.

\noindent {\bf Undressed vertex for two scalar fields and one vector field}.  Using notation as in \rf{man-20092022-01}, we consider cubic vertex for two scalar fields and one vector field,
\beq
&& (\Delta_1,0) - (\Delta_2,0) - (\Delta_3,1)
\nonumber\\
\label{man-21082022-31} && \Delta_a = \frac{d-2}{2} - k \,, \qquad k \in \No_0\,, \qquad a=1,2;\qquad \Delta_3=1\,,
\eeq
i.e. two scalar fields carry external line indices $a=1,2$,
while one vector field carries external line index $a=3$.
For this case, we find one solution  for the undressed vertex given by
\beq
\label{man-21082022-32} && V = \upsilonbf^{\opsm\,\, k} L_3 \,,
\\
\label{man-21082022-33} && \hspace{1cm} 0 \leq k \leq \frac{d-4}{2}\,, \qquad k\in \No_0\,.
\eeq

The following remarks are in order.

\noinbf{i}) Restrictions given in \rf{man-21082022-33}
are obtainable by using the requirements in \rf{man-21082022-09}, \rf{man-21082022-09a}.

\noinbf{ii}) From \rf{man-21082022-31}, we see that the conformal dimensions of two scalar fields coupled to vector field are equal, $\Delta_1=\Delta_2$. Using equation \rf{man-21082022-03} for $a=3$, we verify that, for $\Delta_1\ne \Delta_2$,  conformal invariant cubic interaction of  two scalar fields and one vector field does not exist.

\noinbf{iii}) Restriction for $k$ and $d$ in \rf{man-21082022-33} can be represented as
\be \label{man-21082022-34}
d \geq 2k + 4\,.
\ee
Comparing \rf{man-21082022-26} and \rf{man-21082022-34}, we see that, for the cubic vertex of the three scalar fields, the values of $k_1$, $k_2$, $k_3$ form the upper bound for the allowed values of the space-time dimension $d$, while, for the cubic vertex of the two scalar fields and one vector field, the value of $k$ forms the lower bound for the allowed values of the space-time dimension $d$.

\noinbf{iv}) For $d=4$, $k=0$, we find $UV=V$ and therefore $p_\smp3^- = V$ \rf{man-21082022-01}. For  $d=4$, $k=0$, all fields in \rf{man-21082022-31} have canonical conformal dimension $\Delta= 1$ and these fields are realized as massless fields. Thus, for $d=4$, the vertex $p_\smp3^-=V$ describes the standard cubic interaction of one massless vector field and two massles scalar fields. For $d>4$, we note $UV\ne V$ and therefore $p_\smp3^-\ne V$.

\noinbf{v}) For $d \geq 4$, up to the $\upsilonbf^\opsm$-factor, the undressed vertex \rf{man-21082022-32} coincides with the cubic vertex for two massless scalar fields coupled to one massless vector field in Poincar\'e invariant theory (see, for example, Table I in Ref.\cite{Metsaev:2005ar}%
\footnote{ The variables $L_a$ in this paper are identified with the variables $B_a$ in Ref.\cite{Metsaev:2005ar}.}
).

\noindent {\bf Undressed vertex for two vector fields and one scalar field}.  Using notation as in \rf{man-20092022-01}, we consider cubic vertex for two vector fields and one scalar field,
\beq
&& (\Delta_1,1) - (\Delta_2,1) - (\Delta_3,0)\,,
\nonumber\\
\label{man-21082022-41} && \Delta_1 = 1\,, \qquad \Delta_2 =1\,, \qquad \Delta_3 = 0\,,
\eeq
i.e. two vector fields carry external line indices $a=1,2$,
while one scalar field carries external line index $a=3$.
For this case, we find one solution  for the undressed vertex given by
\be \label{man-21082022-42}
V = \upsilonbf^{\opsm\,\,   \frac{d-4}{2} } L_1 L_2\,.
\ee
\noinbf{i}) As seen from \rf{man-21082022-41}, the conformal dimension of the scalar field coupled to two vector fields is equal to zero, $\Delta_3 = 0$. In Appendix B, we verify that, for $\Delta_3 \ne 0$,  a cubic vertex for one conformal scalar field and two vector fields does not exist (see \rf{man-22082022-01app}-\rf{man-22082022-03app}).

\noinbf{ii}) For $d\geq 4$, we find $UV\ne V$ and therefore $p_\smp3^- \ne V$ \rf{man-21082022-01}. For $d=4$, the vertex $p_\smp3^-$ describes cubic interaction of one scalar field with conformal dimension $\Delta_3=0$ and two massless vector fields. Recall that, for $d=4$, the scalar field with the conformal dimension $\Delta_3=0$ is not realized as a massless field.

\noinbf{iii}) For $d \geq 4$, the undressed vertex \rf{man-21082022-42} coincides, up to the $\upsilonbf^\opsm$-factor, with the cubic vertex for two massless vector fields and one massless scalar field in Poincar\'e invariant theory (see, for example, Table I in Ref.\cite{Metsaev:2005ar}).

\noindent {\bf Undressed vertex for three vector fields}.  Using notation as in \rf{man-20092022-01}, we consider cubic vertex for three vector fields,
\be \label{man-21082022-50}
(\Delta_1,1) - (\Delta_2,1) - (\Delta_3,1)\hspace{2cm} \Delta_a = 1\,, \qquad a=1,2,3\,,
\ee
i.e. three vector fields carry external line indices $a=1,2,3$.
For this case, we find two solutions  for the undressed vertex given by
\beq
\label{man-21082022-51} && V = \upsilonbf^{\opsm\,\, \frac{d-4}{2} } \big( L_1 \alpha_{23} + L_2\alpha_{31} +  L_3\alpha_{12} \big)\,,\hspace{1cm} d\geq 4\,, \qquad d-\even;\qquad\qquad
\\
\label{man-21082022-52} && V = \upsilonbf^{\opsm\,\, \frac{d-6}{2} } L_1 L_2 L_3 \,,\hspace{4.2cm} d\geq 6\,, \qquad d-\even\,.
\eeq

The following remarks are in order.

\noinbf{i}) For $d=4$, we find $UV=V$ and therefore the undressed vertex $V$ \rf{man-21082022-51} is equal to $p_\smp3^-$ \rf{man-21082022-01}. This cubic vertex $p_\smp3^-=V$ describes cubic interaction of three massless vector fields in Yang-Mills theory as it should be. For the undressed vertices \rf{man-21082022-51}, \rf{man-21082022-52} when $d=6$, we note $UV\ne V$ and therefore the undressed vertices \rf{man-21082022-51}, \rf{man-21082022-52} are not equal to the cubic vertices, $p_\smp3^-\ne V$.

\noinbf{ii}) For $d \geq 4$, the two undressed vertices \rf{man-21082022-51}, \rf{man-21082022-52} coincide, up to the $\upsilonbf^\opsm$-factor, with the respective two cubic vertices for three massless vector fields in Poincar\'e invariant theory (see, for example, Table I in Ref.\cite{Metsaev:2005ar}).

We finish our study of the cubic vertices with the following two observations.

\noinbf{a}) For $d\geq 6$, the number of cubic vertices for scalar and vector fields in conformal invariant theory  is equal to the number of cubic vertices for massless scalar and vector fields in Poincar\'e invariant theory.

\noinbf{b}) For $d\geq 6$,  the undressed cubic vertices for scalar and vector fields in conformal invariant theory coincide, up to the $\upsilonbf^\opsm$-factor, with the cubic vertices for massless scalar and vector fields in Poincar\'e invariant theory.

Using these two observations, we now put forward the following two conjectures for arbitrary spin conformal fields propagating in $R^{d-1,1}$, $d\geq 6$.

\noinbf{ Conjecture A}. For $d\geq 6$, the number of cubic vertices for arbitrary spin fields in conformal invariant theory is equal to the number of cubic vertices for massless arbitrary spin fields in Poincar\'e invariant theory. As shown in Ref.\cite{Metsaev:2005ar}, for three massless spin-$s_a$ fields, $a=1,2,3$, the number of the parity-even cubic vertices in Poincar\'e invariant theory is equal to $s_\minrm+1$, where  $s_\minrm = \min_{a=1,2,3} s_a$. Our conjecture implies then that, for three spin-$s_a$ fields, $a=1,2,3$, the number of the parity-even cubic vertices in conformal invariant theory is also equal to $s_\minrm+1$.

\noinbf{Conjecture B}. For $d\geq 6$,  the undressed cubic vertices for arbitrary spin fields in conformal invariant theory coincide, up to the $\upsilonbf^\opsm$-factor, with the cubic vertices for massless arbitrary spin fields in Poincar\'e invariant theory.

Dependence of the cubic vertex $p_\smp3^-$ on the oscillators $\zeta_a$, $\upsilon_a^\omsm$, and $\upsilon_a^\opsm$ (by module of $\upsilonbf^\opsm$) is governed by the dressing operator $U$. To get explicit expression for the cubic vertex  $p_\smp3^-$ in terms of the just mentioned oscillators we should use Taylor series expansion of the dressing operator $U$ in $\partial_{L_a}$ and $\partial_{\alpha_{aa+1}}$. Such expansion turns out to be very complicated and for this reason explicit expression for the cubic vertex $p_\smp3^-$ in terms of the oscillators $\zeta_a$, $\upsilon_a^\omsm$, and $\upsilon_a^\opsm$ is not illuminating.
The possibility to represent the cubic vertex in terms of the undressed vertex and dressing operator \rf{man-21082022-01} we consider as the main attractive feature of our approach. For the reader convenience, we now present the explicit expressions for the cubic vertex $p_\smp3^-$ when the undressed vertex $V$ is a degree-1 polynomial in $L_a$. The corresponding undressed vertices are given in \rf{man-21082022-21}, \rf{man-21082022-32}, \rf{man-21082022-51}.

\noindent {\bf Cubic vertex for three scalar fields}. Using the undressed vertex $V$ \rf{man-21082022-21}, our result for the cubic vertex $p_\smp3^-$ and the corresponding densities can be summarized as,
\beq
\label{man-18122016-02} && p_\smp3^- = \upsilon^n \prod_{a=1,2,3} \upsilon_a^{\ominussm\, n_a} \,, \hspace{1cm} j_\smp3^{-i} = 0 \,, \hspace{1cm} k_\smp3^- = \frac{1}{6} \Mbf^\ommsm p_\smp3^-\,,
\eeq
where $n$, $n_a$ are given in \rf{man-21082022-22}, \rf{man-21082022-22a} and we use the notation for $\Mbf^\ommsm$ given in \rf{man-17122016-25}.

\noindent {\bf Cubic vertices for two scalar fields and one vector field}. Using the undressed vertex $V$ \rf{man-21082022-32}, our result for the cubic vertex $p_\smp3^-$ and the corresponding densities can be summarized as,
\beq
\label{man-18122016-17} p_\smp3^- & = &  L_3 \upsilon^k \upsilon_3^{\omsm\,\, n}
+ \frac{k}{\beta_3}\zeta_3 \upsilon_{12} \upsilon^{k-1}  \upsilon_3^{\omsm\,\, n}
- \frac{n \betach_3}{2\beta_3} \zeta_3  \upsilon^k   \upsilon_3^{\omsm\,\, n -1}\,,\qquad
\nonumber\\
\label{man-18122016-18} j_\smp3^{-i}  & = &  - \frac{2\betach_3}{3\beta_3}  \alpha_3^i \upsilon^k \upsilon_3^{\omsm\,\, n}\,,
\nonumber\\
\label{man-18122016-19} k_\smp3^-  & = &  - \frac{4k\betach_3}{3\beta_3} \zeta_3 \upsilon_1^\omsm \upsilon_2^\omsm \upsilon^{k-1} \upsilon_3^{\omsm\,\, n} + \frac{1}{6} \Mbf^\ommsm p_\smp3^-\,, \qquad n \equiv \frac{d-4}{2} - k\,, \hspace{1cm}
\\
\label{man-18122016-20} && \upsilon_{ab} \equiv \upsilon_a^\opsm \upsilon_b^\omsm \beta_b  - \upsilon_b^\opsm \upsilon_a^\omsm \beta_a \,,\qquad
\eeq
where $n\geq 0$. The $L_a$ and $\upsilon$ are defined in \rf{man-16082022-03} and \rf{man-21082022-18}, while $\Mbf^\ommsm$ is defined in \rf{man-17122016-25}.

\noindent {\bf Cubic vertex for three vector fields}.
Using the undressed vertex $V$ \rf{man-21082022-51}, our result for the cubic vertex $p_\smp3^-$ and the corresponding densities can be summarized as,
\beq
\label{man-18122016-26} p_\smp3^- & = & \sum_{a=1,2,3} L_a \big( v \alpha_{a+1a+2} - m \zeta_{a+1}\zeta_{a+2} \upsilon_a^\omsm \big) \upsilon^{m-1}
\nonumber\\
& +  & \sum_{a=1,2,3} \frac{m\zeta_a}{\beta_a} \big( v \alpha_{a+1a+2} - (m-1) \zeta_{a+1}\zeta_{a+2} \upsilon_a^\omsm \big) v_{a+1a+2} \upsilon^{m-2}
\nonumber\\
& - &  \frac{m\check\beta}{2\beta} \zeta_1\zeta_2\zeta_3 \upsilon^{m-1}\,,
\nonumber\\
\label{man-18122016-27}  j_\smp3^{-i} &  =  & - \sum_{a=1,2,3} \frac{2\check\beta_a}{3\beta_a}  \alpha_a^i \big(v \alpha_{a+1a+2} - m \zeta_{a+1}\zeta_{a+2} \upsilon_a^\omsm \big) \upsilon^{m-1}\,,
\nonumber\\
\label{man-18122016-28} k_\smp3^-  & = & - \sum_{a=1,2,3} \frac{4m\check\beta_a\zeta_a}{3\beta_a}  \upsilon_{a+1}^\omsm \upsilon_{a+2}^\omsm \big(v \alpha_{a+1a+2} - (m -1) \zeta_{a+1}\zeta_{a+2} \upsilon_a^\omsm \big) \upsilon^{m-2}
\nonumber\\
& + & \frac{1}{6} \Mbf^\ommsm p_\smp3^-\,, \hspace{1cm} m\equiv \frac{d-4}{2}\,, %
\eeq
where $\Mbf^\ommsm$, $L_a$, and $\alpha_{ab}$  are given in \rf{man-17122016-25}, \rf{man-16082022-03}, while $\upsilon$ and $\upsilon_{ab}$ are given in \rf{man-21082022-18}, \rf{man-18122016-20}.

\noinbf{Incorporation of internal $o(\Nsf)$ symmetry}.%
\footnote{ Incorporation of internal $u(\Nsf)$ and $usp(\Nsf)$ symmetries in higher-spin theories can be found in Refs.\cite{Skvortsov:2020wtf}.}
We now demonstrate a procedure of the incorporation of internal $o(\Nsf)$ symmetry. Let us use the shortcut $\phi_{\Delta,s}^\Ism$ for the component fields (scalars, vectors, and tensors) entering the ket-vectors in \rf{man-14122016-10}, \rf{man-14122016-15}, \rf{man-14122016-27}.  In place of a singlet conformal field $\phi_{\Delta,s}^\Ism$, we introduce colored conformal fields $\phi_{\Delta,s}^{\Ism\,\,\asf\bsf}$, where the matrix indices of the $o(\Nsf)$ algebra take values $\asf,\bsf = 1,\ldots,\Nsf$.  By definition, the colored conformal fields satisfy the relations
\beq
&& \phi_{\Delta,s}^{\Ism\,\,\asf\bsf}(x^+,p) = (-)^s \phi_{\Delta,s}^{\Ism\,\,\bsf\asf}(x^+,p)\,, \qquad (\phi_{\Delta,s}^{\Ism\,\,\asf\bsf}(x^+,p))^\dagger = \phi_{\Delta,s}^{\Ism\,\,\asf\bsf}(x^+,-p)\,.
\eeq
The products of singlet ket-vectors in \rf{man-15122016-01}, \rf{man-15122016-04} and \rf{man-15122016-09} should be replaced as
\be
\langle\phi_{\Delta,s}|\phi_{\Delta,s}\rangle \rightarrow \langle \phi_{\Delta,s}^{\asf\bsf} |\phi_{\Delta,s}^{\asf\bsf}\rangle\,,\qquad
\prod_{a=1,2,3}\langle\phi_{\Delta_a,s_a}| \rightarrow
\langle\phi_{\Delta_1,s_1}^{\asf\bsf}| \langle\phi_{\Delta_2,s_2}^{\bsf\csf}| \langle\phi_{\Delta_3,s_3}^{\csf\asf}|\,,
\ee
where $\langle\phi_{\Delta,s}^{\asf\bsf}|\equiv |\phi_{\Delta,s}^{\asf\bsf}\rangle{}^\dagger$, while the equal-time commutator \rf{man-15122016-02} should be replaced as
\beq
&&  \hspace{-0.8cm} [|\phi_{\Delta,s}^{\asf\bsf}(x^+,p)\rangle, |\phi_{\Delta',s'}^{\asf'\bsf'}(x^+,p^\prime)\rangle\,] =  \frac{1}{2\beta} \delta(\beta+\beta^\prime)\delta^{d-2}(p+p^\prime)
|\rangle |\rangle' \Pi_s^{\asf\bsf,\asf'\bsf'} \delta_{\Delta,\Delta'} \delta_{s,s'}\,,\qquad
\\
&& \hspace{-0.8cm} \Pi_s^{\asf\bsf,\asf'\bsf'} \equiv \half\big( \delta^{\asf\asf'} \delta^{\bsf\bsf'} + (-)^s \delta^{\asf\bsf'} \delta^{\bsf\asf'} \big)\,, \qquad \Pi_s^{\asf\bsf,\asf'\bsf'} \Pi_s^{\asf'\bsf',\csf\esf} = \Pi_s^{\asf\bsf,\csf\esf}\,.
\eeq
The $o(\Nsf)$ algebra generators denoted as $J^{\asf\bsf}$,  $J^{\asf\bsf}= - J^{\bsf\asf}$, are realized as
\be
J^{\asf\bsf} =  2 \int d^{d-1}p\, \beta \big( \langle\phi_{\Delta,s}^{\asf\csf} |\phi_{\Delta,s}^{\bsf\csf}\rangle - \langle\phi_{\Delta,s}^{\bsf\csf} |\phi_{\Delta,s}^{\asf\csf}\rangle \big)\,.
\ee
The generators $J^{\asf\bsf}$ and conformal fields $|\phi_{\Delta,s}^{\asf\bsf}\rangle$ obey the commutators
\be
[J^{\asf\bsf},J^{\csf\esf} ] = \delta^{\bsf\csf} J^{\asf\esf} + 3 \hbox{ terms}, \qquad [|\phi_{\Delta,s}^{\asf\bsf}\rangle,J^{\csf\esf} ] = \delta^{\bsf\csf} |\phi_{\Delta,s}^{\asf\esf}\rangle + 3 \hbox{ terms}.
\ee

\newsection{ \large Conclusions  }\label{seccc-07}

In this paper, by using the ordinary-derivative light-cone gauge approach, we considered interacting arbitrary spin totally symmetric conformal fields propagating in the $R^{d-1,1}$ space, $d\geq 4$.
For such fields, we obtained restrictions imposed on the cubic vertices by kinematical and dynamical symmetries of the conformal algebra $so(d,2)$.

We used these restrictions for the detailed study of interacting scalar and vector fields  and
  found all parity-even cubic vertices. We presented also all restrictions on the allowed values of conformal dimensions of the scalar and vectors fields entering our cubic vertices. Using our results for the scalar and vector fields, we put forward a conjecture about cubic vertices for the arbitrary spin conformal fields propagating in the $R^{d-1,1}$ space.

  Namely, for $d\geq 6$, we expect
that, given conformal fields with spin values $s_1$, $s_2$, $s_3$, the number of parity-even light-cone gauge cubic vertices that can be built in conformal invariant theory is equal to  $s_\minrm+1$, where $ s_\minrm \equiv \min_{a=1,2,3} s_a$. We, recall that, according to the result in Ref.\cite{Metsaev:2005ar},
given massless fields with spin values $s_1$, $s_2$, $s_3$, the number of parity-even light-cone gauge cubic vertices that can be built in Poincar\'e invariant theory is also equal to $s_\minrm+1$.%
\footnote{ The corresponding Lorentz covariant cubic vertices by using various approaches were built in Refs.\cite{Manvelyan:2010jr}-\cite{Buchbinder:2021xbk}.}

Also we expect that the undressed cubic vertex for conformal fields in conformal invariant theory coincides, up to some factor, with cubic vertex of massless fields in Poincar\'e invariant theory. For $d\geq 6$, in Poincar\'e invariant theories, the number of parity-even Lorentz covariant cubic vertices is equal to the number of parity-even light-cone gauge cubic vertices. Conjecturing that the same equality holds true in conformal invariant theory, we note then that,
given conformal fields with spin values $s_1$, $s_2$, $s_3$, the number of Lorentz covariant parity-even cubic vertices in conformal invariant theory is equal to $s_\minrm+1$.

We now mention the following applications and generalizations of our results.

\noinbf{i}) Interesting application of our formalism is related to light-cone gauge cubic vertices for conformal graviton in the $R^{d-1,1}$ space for $d\geq 6$. For   $d=6$ and $d=8$, the Lorentz covariant cubic vertices for conformal graviton are available from the literature (see Refs.\cite{Bonora:1985cq,Boulanger:2004zf}). However to our knowledge, for arbitrary $d$, Lorentz covariant as well as light-cone gauge cubic vertices for conformal graviton are not available in the literature so far. For arbitrary $d\geq 6$, our conjecture implies that for conformal graviton there is only three light-cone gauge cubic vertices. Note that for the counting of light-cone gauge cubic vertices and their Lorentz covariant cousins we use only those vertices which lead to nontrivial and different 3-point amplitudes.

For $d=6$, there are three local Weyl invariants in Ref.\cite{Bonora:1985cq}. In view of our conjecture we expect that all of them lead to nontrivial and different 3-point amplitudes. For $d=8$, there are five local Weyl invariants in Ref.\cite{Boulanger:2004zf}.%
\footnote{ These 5 invariants are of the type $D^4 C^2 + D^2 C^3$, where $D$ and $C$ stand for the covariant derivative and the Weyl tensor respectively. There are also 7 invariants of the type $C^4$ (see Ref.\cite{Fulling:1992vm}). These 7 invariants do not contribute to cubic amplitudes and therefore are not relevant for our discussion.}
In view of our conjecture, we expect that only three of those five invariants (or three their combinations) lead to nontrivial and different 3-point amplitudes.
For the reader convenience, we recall that, as noted in Ref.\cite{Metsaev:1986yb}, in Poincar\'e invariant theory,  for graviton field in dimensions of greater or equal to six there are also only three Lorentz covariant cubic vertices which lead to nontrivial and different 3-point amplitudes. The corresponding three light-cone gauge cubic vertices for graviton field were worked out in Ref.\cite{Metsaev:2005ar}.

As a side remark we note that a number of cubic vertices for conformal fields can in principle be understood by using the AdS/CFT correspondence. We recall that an action of a AdS field evaluated on a solution of the Dirichlet problem is referred to as effective action. As shown in Ref.\cite{Liu:1998bu}, the UV divergence of the effective action for the AdS Einstein graviton field is realized as the Weyl action for the conformal graviton field.
For the graviton field in the $AdS_{d+1}$ space, $d\geq 6$, there are three cubic vertices. Therefore one can expect that the UV divergence of the effective action for those three cubic vertices of the AdS graviton field is realized as three cubic vertices for the conformal graviton field.%
\footnote{ As shown in Ref.\cite{Metsaev:2009ym}, the UV divergence of the effective action for free higher-spin AdS fields is realized as action for free higher-spin conformal fields. For arbitrary spin fields in $AdS_{d+1}$, $d\geq 6$, there are $s_\minrm+1$ cubic vertices (see Ref.\cite{Joung:2011ww}). We expect then that the UV divergence of the effective action for those $s_\minrm+1$ cubic vertices of AdS fields is realized as $s_\minrm+1$ cubic vertices for conformal fields.
}
Recall that, in CFT, the 3-point correlator of the energy-momentum tensor is decomposed, in general, in the three tensor structures, when $d\geq 6$.

\noinbf{ii}) As is well known, the parity-even light-cone gauge cubic vertices can relatively straightforwardly be uplifted to their BRST-BV cousins (see, e.g., Refs.\cite{Siegel:1986zi,Metsaev:2012uy}). Ordinary-derivative BRST-BV formulation of free conformal fields has already been developed in Ref.\cite{Metsaev:2015yyv}. We expect therefore that the method of the dressing operators and undressed vertices we developed in this paper can straightforwardly be generalized to BRST-BV interacting conformal fields.

\noinbf{iii}) Various methods for the study of light-cone gauge arbitrary spin field theories invariant with respect to the Poincar\'e supersymmetries were discussed in Refs.\cite{Bengtsson:1983pg}-\cite{Metsaev:2019aig}. We expect that results in this paper and the ones in Refs. \cite{Bengtsson:1983pg}-\cite{Metsaev:2019aig}  provide a good starting point for study of supersymmetric light-cone gauge conformal fields. Application of our light-cone approach along the lines in Refs.\cite{Beccaria:2017nco}-\cite{Kuzenko:2022hdv} could also be of some interest.

\medskip

{\bf Acknowledgments}. This work was supported by the RFBR Grant No.20-02-00193.

\setcounter{section}{0}\setcounter{subsection}{0}
\appendix{ \large Prove of Statement on vertices \rf{man-16082022-02}, \rf{man-16082022-04} }

Following presentation in Appendix B in Ref.\cite{Metsaev:2005ar}, we note that under a field redefinition governed by a density $f$ the vertex $p_\smp3^-$ \rf{man-16082022-02} is changed as
\beq
\label{man-23082022-01} && p_\smp3^-{}^f = p_\smp3^- + \Pbf^- f\,,
\hspace{1.2cm} f =  f(\Po^i, \alpha_a^i, \zeta_a\,, \upsilon_a^\omsm\,, \upsilon_a^\opsm\,, \beta_a)\,,
\eeq
where the density $f$ satisfies the kinematical equations,
\beq
\label{man-23082022-02} && (N_\Po + \Nbf_\beta -1\big) f = 0\,, \hspace{1cm} \big( N_{\Po}  + \frac{d-2}{2} -  \Mbf^\ompsm \big) f  = 0 \,,\qquad
\\
\label{man-23082022-04} && \Jbf^{ij} f = 0 \,,
\\
\label{man-23082022-05} && \Kbf^+ f =  0 \,, \qquad  \Kbf^i f =  0 \,,
\eeq
and our conventions are summarized in \rf{man-17122016-17}-\rf{man-17122016-27}. We recall also that the vertex $p_\smp3^-$ satisfies kinematical equations \rf{man-17122016-05}-\rf{man-17122016-28} and dynamical equation \rf{man-17122016-41}. We now prove the Statement in \rf{man-16082022-04} which tells us that the density $f$ can be chosen so that the vertex $p_\smp3^{-f}$ becomes independent of $\Po^2$. We prove the Statement  in  the following four steps.

\noinbf{ Step 1}. In view of the $so(d-2)$ symmetries \rf{man-23082022-04}, the density $f$ can be presented as
\be \label{man-23082022-07}
f =  f(L_a, \alpha_{aa+1}, \zeta_a\,, \upsilon_a^\omsm\,, \upsilon_a^\opsm\,, \beta_a\,, \Po^2)\,.
\ee

\noinbf{ Step 2}. By definition, the vertex $p_\smp3^-$ and the density $f$ are finite order polynomials of $\Po^i$. Therefore the vertex $p_\smp3^-$ and the density $f$ can be expanded in the Taylor series in $\Po^2$. In view of \rf{man-17122016-39}, we note that the Taylor series expansion in $\Po^2$ can be represented as the Taylor series expansion in $\Pbf^-$. This is to say that the vertex $p_\smp3^-$ and the density $f$ can be presented as
\beq
\label{man-23082022-08} && p_\smp3^- = \sum_{n=0}^N (\Pbf^-)^n  V_n\,, \hspace{1.2cm} V_n =  V_n(L_a, \alpha_{aa+1}, \zeta_a\,, \upsilon_a^\omsm\,, \upsilon_a^\opsm\,, \beta_a)\,,
\\
\label{man-23082022-09} && f  = \sum_{n=1}^N (\Pbf^-)^{n-1}  f_n\,,\hspace{1cm} f_n =  f_n(L_a, \alpha_{aa+1}, \zeta_a\,, \upsilon_a^\omsm\,, \upsilon_a^\opsm\,, \beta_a)\,.
\eeq
Using \rf{man-23082022-01}, \rf{man-23082022-08}, and \rf{man-23082022-09}, we find
\be \label{man-23082022-10}
p_\smp3^-{}^f = V_0 + \sum_{n=1}^N (\Pbf^-)^n  (V_n + f_n)\,.
\ee
From \rf{man-23082022-10}, we see that the choice
\be \label{man-23082022-11}
f_n = - V_n\,, \qquad n=1,\ldots, N\,,
\ee
allows us to cast the $p_\smp3^-{}^f$ into the desired form given in \rf{man-16082022-04}. Note however that the choice for $f_n$ \rf{man-23082022-11} is possible only if the densities $f_n$ satisfy the same equations as the vertices $V_n$. Thus all that remains is to prove that $f_n$ and  $V_n$, $n=1,\ldots,N$, satisfy one and same equations.

\noinbf{ Step 3}. Using notation in \rf{man-17122016-26}, \rf{man-17122016-26a}, we find that the equations for $p_\smp3^-$ in \rf{man-17122016-05} and \rf{man-17122016-08} lead to the following equations for $V_n$, $n=0,1,\ldots,N$,
\be  \label{man-23082022-12}
\big( \Nbf_L  + \Nbf_\beta - n \big) V_n  = 0 \,, \qquad \big( \Nbf_L  + \frac{d-6}{2} -  \Mbf^\ompsm + 2n \big) V_n  = 0 \,,
\ee
while equations \rf{man-23082022-02} lead to equations for $f_n$ which are obtained from equations  \rf{man-23082022-12} by using the substitution $V_n\rightarrow f_n$, $n=1,\ldots,N$. This implies that equations \rf{man-17122016-05}, \rf{man-17122016-08} and \rf{man-23082022-02} lead to one and same equations for the respective $V_n$ and $f_n$, $n=1,\ldots,N$.

\noinbf{Step 4}. Using the commutators for the operators defined in \rf{man-17122016-18}-\rf{man-17122016-40},
\beq
\label{man-23082022-14} && [\Kbf^+,\Pbf^-]  =    N_\Po + \frac{d-2}{2} - \Mbf^\ompsm\,, \hspace{1cm} [\Kbf^+, \Jbf^{-i}]   =    - \Kbf^i\,,
\\
\label{man-23082022-16} && [\Kbf^i, \Pbf^-] = - \Jbf^{-i}\,, \hspace{3.7cm} [\Pbf^-,\Jbf^{-i}]= 0\,,
\eeq
and equations \rf{man-23082022-12}, we find the following relations:
\beq
\label{man-23082022-17} && \Kbf^+ p_\smp3^-   =    \Kbf^+ V_0 + \Pbf^- L^+(V_1,\ldots,V_N)\,,
\\
\label{man-23082022-18} && L^+(V_1,\ldots,V_N) \equiv \sum_{n=1}^{N-1} (\Pbf^-)^{n-1} \big(\Kbf^+  V_n - n(n+1) V_{n+1}\big) +  (\Pbf^-)^{N-1} \Kbf^+ V_N\,, \qquad \qquad
\\
\label{man-23082022-19} && (\Jbf^{-i} + \Pbf^- \Kbf^i) p_\smp3^-  = \bigl( \Jbf^{-i}   + \Pbf^- \Kbf^i \bigr) V_0 + \Pbf^-\Pbf^- L^i(V_1,\ldots,V_N)\,,
\\
\label{man-23082022-20} && L^i(V_1,\ldots,V_N) \equiv \sum_{n=2}^N (\Pbf^-)^{n-2} \Bigl( \Kbf^i V_{n-1} - (n-1) \Jbf^{-i} V_n\Bigr) +   (\Pbf^-)^{N-1} \Kbf^i V_N\,. \qquad \qquad
\eeq
Representation of the operator $\Kbf^+$ on vertices $V_n$, $n=0,1,\ldots,N$, given in \rf{man-16082022-05} implies that the action of the operator $\Kbf^+$ on the vertex $V_0$ does not produce $\Pbf^-$-terms. In turn, this implies that equation for the cubic vertex $p_\smp3^-$ in \rf{man-17122016-28} and relation \rf{man-23082022-17} amount to the following equations:
\beq
\label{man-23082022-21} && \Kbf^+ V_0 = 0 \,,
\\
\label{man-23082022-22} && L^+(V_1,\ldots,V_N) = 0 \,.
\eeq
Representation of the operators $\Kbf^i$ and $\Jbf^{-i}$ on vertices $V_n$, $n=0,1,\ldots,N$, given in \rf{man-16082022-06}, \rf{man-16082022-07} implies that the action of the operator $\Kbf^i$ on the vertex $V_0$ does not produce $\Pbf^-$-terms, while the action of the operator $\Jbf^{-i}$ on the vertex $V_0$ does not produce terms higher than first order in $\Pbf^-$. In turn, this implies that equation \rf{man-17122016-41} and relation \rf{man-23082022-19} amount to the following equations:
\beq
\label{man-23082022-23} && \bigl( \Jbf^{-i}   + \Pbf^- \Kbf^i \bigr) V_0 =0\,,
\\
\label{man-23082022-24} &&  L^i(V_1,\ldots,V_N) = 0 \,.
\eeq
We now consider equations for $f$ in \rf{man-23082022-05}. Straightforward computation gives the relations
\be \label{man-23082022-25}
\Kbf^+ f  = L^+(f_1,\ldots,f_N)\,, \qquad \Kbf^i f  =   L^i(f_1,\ldots,f_N)\,,
\ee
where $L^+(f_1,\ldots,f_N)$ and $L^i(f_1,\ldots,f_N)$ are obtained from \rf{man-23082022-18} and \rf{man-23082022-20} respectively  by the substitution $V_n\rightarrow f_n$. Equations \rf{man-23082022-05} and relations \rf{man-23082022-25} give the equations
\be \label{man-23082022-26}
L^+(f_1,\ldots,f_N) = 0 \,, \qquad  L^i(f_1,\ldots,f_N) = 0\,.
\ee
Comparing \rf{man-23082022-22}, \rf{man-23082022-24} and \rf{man-23082022-26}, we conclude that $V_n$ and $f_n$, $n=1, \ldots, N$, satisfy one and same equations. Hence we can use fields redefinition with $f_n$ as in \rf{man-23082022-11} and this gives the desired representative $V_0$. In Secs. \ref{seccc-05}, \ref{seccc-06}, this $V_0$ is denoted simply as $p_\smp3^-$ \rf{man-16082022-04}.

\appendix{ \large Method for derivation of cubic vertex  \rf{man-21082022-01} }

In this Appendix,  given operators $A$ and $X$, the operator $A|_X$ is defined as
\be \label{man-22082022-00}
A|_X \equiv X^{-1} A X \,.
\ee
For the derivation of vertex $p_\smp3^-$ \rf{man-21082022-01} and equations \rf{man-21082022-03}-\rf{man-21082022-05} we use the equations given in \rf{man-16082022-31}-\rf{man-16082022-37}. We split our derivation in six steps.

\noinbf{Step 1}. We consider equations \rf{man-16082022-31}, \rf{man-16082022-34}. Using operators $U_\beta$ \rf{man-21082022-11} and  $G_\beta^\Jbf$ \rf{man-16082022-23}, we introduce the vertex $V^\smone$ and the operator $G_\beta^{\Jbf\smone}$,
\be \label{man-22082022-01}
p_\smp3^- = U_\beta V^\smone\,, \qquad G_\beta^{\Jbf\smone} \equiv G_\beta^\Jbf|_{U_\beta}\,,\qquad G_\beta^{\Jbf\smone}  =  - \beta^{-1}  \No_\beta \,.
\ee
In terms of the vertex $V^\smone$ \rf{man-22082022-01}, equation \rf{man-16082022-31} takes the form $\Nbf_\beta V^\smone=0$, while equation \rf{man-16082022-34} takes the form $G_\beta^{\Jbf\smone} V^\smone = 0$, i.e., $\No_\beta  V^\smone = 0$. Thus, we obtain two equations
\be \label{man-22082022-03}
\Nbf_\beta V^\smone = 0 \,, \qquad \No_\beta  V^\smone = 0\,.
\ee
Equations \rf{man-22082022-03} tell us that the vertex $V^\smone$ is independent of the momenta $\beta_1$, $\beta_2$, $\beta_3$,
\be \label{man-22082022-05}
V^\smone = V^\smone (L_a, \alpha_{aa+1}, \zeta_a, \upsilon_a^\omsm, \upsilon_a^\opsm)\,.
\ee

\noinbf{Step 2}. We now use equations \rf{man-16082022-39} to fix a dependence of the vertex $V^\smone$ \rf{man-22082022-05} on the oscillators $\upsilon_1^\omsm$, $\upsilon_2^\omsm$, $\upsilon_3^\omsm$. Namely, by using operator $U_{\upsilon^\omsm}$ \rf{man-21082022-12}, the equations \rf{man-16082022-39} can be solved as
\be \label{man-22082022-06}
V^\smone = U_{\upsilon^\omsm} V^\smtwo\,, \qquad  V^\smtwo =V^\smtwo(L_a, \alpha_{aa+1}, \zeta_a, \upsilon_a^\opsm)\,,
\ee
where the vertex $V^\smtwo$ is independent of the oscillators $\upsilon_1^\omsm$, $\upsilon_2^\omsm$, $\upsilon_3^\omsm$.

\noinbf{Step 3}. We consider equations \rf{man-16082022-33}, \rf{man-16082022-35}. First, we introduce the operators $\Kbf^{+\smone}$, $G_\beta^{\Kbf\smone}$,
\beq
\label{man-22082022-08} && \Kbf^{+\smone} \equiv  \Kbf^+\big|_{U_\beta} \,, \hspace{1.5cm} \Kbf^{+\smone}  = \half \sum_{a=1,2,3}   \beta_a K_a^{\ommsm \smone} \,,
\nonumber\\
\label{man-22082022-09}  &&  G_\beta^{\Kbf\smone} \equiv  G_\beta^\Kbf\big|_{U_\beta}\,, \hspace{1.5cm} G_\beta^{\Kbf\smone} =     -  \sum_{a=1,2,3}  \frac{\beta_a \betach_a}{6\beta}  K_a^{\ommsm\smone}\,,
\nonumber\\
\label{man-22082022-11} && K_a^{\ommsm\smone} \equiv  M_a^\ommsm + m_{a+1}^\omsm \partial_{L_{a+1}}  - m_{a+2}^\omsm \partial_{L_{a+2}} + 2 \alpha_{a+1a+2}\partial_{L_{a+1}} \partial_{L_{a+2}}\,.
\eeq
In terms of the vertex $V^\smone$ \rf{man-22082022-01}, \rf{man-22082022-05} equations \rf{man-16082022-33}, \rf{man-16082022-35} take the form
\be \label{man-22082022-12}
\Kbf^{+\smone} V^\smone  = 0\,, \qquad G_\beta^{\Kbf\smone} V^\smone  = 0\,.
\ee
Using the 2nd relation in \rf{man-17122016-01}, we verify that equations \rf{man-22082022-12} amount to the following equations:
\be \label{man-22082022-14}
K_1^{\ommsm\smone}V^\smone = K_2^{\ommsm\smone}V^\smone = K_3^{\ommsm\smone}V^\smone\,.
\ee
Now we represent equations \rf{man-22082022-14} in terms of the vertex $V^\smtwo$. We introduce the operators $K_a^{\ommsm\smtwo}$,
\beq
\label{man-22082022-15} && K_a^{\ommsm\smtwo} \equiv    K_a^{\ommsm\smone}\big|_{U_{\upsilon^\omsm}}\,,
\\
\label{man-22082022-16} && K_a^{\ommsm\smtwo} =  4\partial_{\upsilon_a^\opsm} + m_{a+1}^\omsm \partial_{L_{a+1}}  - m_{a+2}^\omsm \partial_{L_{a+2}} + 2 \alpha_{a+1a+2}\partial_{L_{a+1}} \partial_{L_{a+2}}\,,
\eeq
and note that, in terms of the vertex $V^\smtwo$ \rf{man-22082022-06}, equations \rf{man-22082022-14} take the form
\beq
\label{man-22082022-17}  &&  K_1^{\ommsm\smtwo} V^\smtwo = K_2^{\ommsm\smtwo} V^\smtwo =  K_3^{\ommsm\smtwo} V^\smtwo \,.
\eeq
Solution to equations \rf{man-22082022-17} is found to be
\be \label{man-22082022-18}
V^\smtwo = U_{\upsilon^\opsm}  V^\smthree \,, \qquad  V^\smthree =V^\smthree(L_a, \alpha_{aa+1}, \zeta_a, \upsilonbf_\Lsm^\opsm)\,,
\qquad  \upsilonbf_\Lsm^\opsm  \equiv  \sum_{a=1,2,3} r_a \upsilon_a^\opsm\,,
\ee
where $U_{\upsilon^\opsm}$ and $r_a$ are given in \rf{man-21082022-13}, \rf{man-21082022-14}. In view of $r_a$, the $\upsilonbf_\Lsm^\opsm$ \rf{man-22082022-18} is operator-valued. To avoid the use of the operator-valued variable we introduce the operator $U_{N_\upsilon^\opsm}$ \rf{man-21082022-14} and the vertex $V^\smfour$,
\be \label{man-22082022-21}
V^\smthree = U_{N_{\upsilon^\opsm}}  V^\smfour \,, \qquad V^\smfour =V^\smfour(L_a, \alpha_{aa+1}, \zeta_a, \upsilonbf^\opsm)\,.
\ee

\noinbf{Step 4}. We consider equations \rf{man-16082022-36}. In terms of the vertex $V^\smone$ \rf{man-22082022-01}, equations \rf{man-16082022-36} take the form
\beq
\label{man-22082022-23} && G_a^{\Jbf\Kbf\smone} V^\smone = 0 \,, \hspace{1cm}  G_a^{\Jbf\Kbf\smone} \equiv  G_a^{\Jbf\Kbf}\big|_{U_\beta}\,,\qquad a=1,2,3\,,
\\
\label{man-22082022-24} && G_a^{\Jbf\Kbf\smone}    =  \frac{1}{\beta_a}   \No_\beta\partial_{L_a}
+ \frac{\betach_a}{6\beta_a} K_a^{\ompsm\smone} + K_a^{\omsm\smone}\,,
\\
\label{man-22082022-25} && K_a^{\ompsm\smone}  =  \Big( \Mbf^\ompsm -  N_{L_{a+1}} - N_{L_{a+2}} - \frac{d-4}{2} \Big) \partial_{L_a}\,,
\\
\label{man-22082022-26} && K_a^{\omsm\smone} =  \mb_a^\omsm +  m_{a+1}^\omsm \partial_{\alpha_{aa+1}} +  m_{a+2}^\omsm \partial_{\alpha_{a+2a}} +  \alpha_{a+1 a+2}  \big( \partial_{\alpha_{aa+1}} \partial_{L_{a+2}} -  \partial_{\alpha_{a+2a}} \partial_{L_{a+1}} \big)\qquad\qquad
\nonumber\\
&& \hspace{1cm} +\,\,  \half  \Big(M_{a+1}^\ompsm - M_{a+2}^\ompsm + N_{L_{a+1}} - N_{L_{a+2}} + m_{a+1}^\opsm \partial_{L_{a+1}} + m_{a+2}^\opsm \partial_{L_{a+2}} \Big) \partial_{L_a}\,.
\eeq
Using the 2nd equation in \rf{man-22082022-03}, we note that,  in equation \rf{man-22082022-23}, the $\No_\beta$-term cancels.
To see cancellation of the $K_a^{\ompsm\smone}$-term in equation \rf{man-22082022-23} we note that, in terms of the vertex $V^\smone$ \rf{man-22082022-01}, equation \rf{man-16082022-32} takes the form
\be \label{man-22082022-26a}
\bigl( \Nbf_L  + \frac{d-6}{2} -  \Mbf^\ompsm \bigr) V^\smone  = 0 \,.
\ee
In view of equation \rf{man-22082022-26a} and the relation
\be \label{man-22082022-26b}
K_a^{\ompsm\smone} V^\smone = - \partial_{L_a} \bigl( \Nbf_L  + \frac{d-6}{2} -  \Mbf^\ompsm \bigr)V^\smone\,,
\ee
we see that the $K_a^{\ompsm\smone}$-term in \rf{man-22082022-23} is indeed cancelled. Thus equations \rf{man-22082022-23} amount to equations
\be \label{man-22082022-27}
K_a^{\omsm\smone} V^\smone = 0 \,, \qquad a=1,2,3\,.
\ee
Now we should reformulate equations \rf{man-22082022-27} in terms of the vertex  $V^\smfour$ defined in \rf{man-22082022-06}, \rf{man-22082022-18}, \rf{man-22082022-21}. To this end we note that,
in terms of the vertex $V^\smfour$, equations \rf{man-22082022-27} take the form
\be \label{man-22082022-28}
K_a^{\omsm\smfour} V^\smfour = 0 \,, \qquad a=1,2,3\,,
\ee
where the operators $K_a^{\omsm\smfour}$ are obtained from $K_a^{\omsm\smone}$ by using the following sequence of transformations:
\be \label{man-22082022-29}
K_a^{\omsm\smfour} = K_a^{\omsm\smthree}\big|_{U_{N_{\upsilon^\opsm}}}\,, \qquad K_a^{\omsm\smthree} = K_a^{\omsm\smtwo}|_{U_{\upsilon^\opsm}}\,,
\qquad K_a^{\omsm\smtwo} = K_a^{\omsm\smone}\big|_{U_{\upsilon^\omsm}}\,.
\ee
Our result for the operators defined in \rf{man-22082022-29} is given by
\beq
\label{man-22082022-30} && K_a^{\omsm\smtwo}  =  -2\partial_{\zeta_a} +  \alpha_{a+1 a+2}  \big( \partial_{\alpha_{aa+1}} \partial_{L_{a+2}} -  \partial_{\alpha_{a+2a}} \partial_{L_{a+1}} \big)
\nonumber\\
&&\hspace{1cm}  +\,\,  \half  \Big( \kb_{s_{a+1}} - \kb_{s_{a+2}} - 2 N_{\upsilon_{a+1}^\opsm} + 2 N_{\upsilon_{a+2}^\opsm} + 2N_{L_{a+1}} - 2N_{L_{a+2}} \Big) \partial_{L_a}
\nonumber\\
&& \hspace{1cm} +\,\, \half  \Big( \zeta_{a+1}(\frac{d-4}{2} - N_{\upsilon_{a+1}^\opsm})\partial_{L_{a+1}} + \zeta_{a+2} (\frac{d-4}{2} - N_{\upsilon_{a+2}^\opsm}) \partial_{L_{a+2}} \Big) \partial_{L_a} \qquad
\nonumber\\
&& \hspace{1cm} +\,\,   m_{a+1}^\omsm \partial_{\alpha_{aa+1}} +  m_{a+2}^\omsm \partial_{\alpha_{a+2a}} \,.
\\
\label{man-22082022-31} && K_a^{\omsm\smthree}  =  -2\partial_{\zeta_a} +  \alpha_{a+1 a+2}  \big( \partial_{\alpha_{aa+1}} \partial_{L_{a+2}} -  \partial_{\alpha_{a+2a}} \partial_{L_{a+1}} \big)
\nonumber\\
&& \hspace{1cm}  +\,\, \half  \Big( \kb_{s_{a+1}} - \kb_{s_{a+2}} - 2 N_{\upsilon_{a+1}^\opsm} + 2 N_{\upsilon_{a+2}^\opsm} + 2N_{L_{a+1}} - 2N_{L_{a+2}} \Big) \partial_{L_a}
\nonumber\\
&&\hspace{1cm}  +\,\, \half  \Big( \zeta_{a+1}(\frac{d-4}{2} - N_{\upsilon_{a+1}^\opsm})\partial_{L_{a+1}} + \zeta_{a+2} (\frac{d-4}{2} - N_{\upsilon_{a+2}^\opsm}) \partial_{L_{a+2}} \Big) \partial_{L_a} \qquad
\nonumber\\
&& \hspace{1cm}  +\,\,  m_{a+1}^\omsm \partial_{\alpha_{aa+1}} +  m_{a+2}^\omsm \partial_{\alpha_{a+2a}}
\nonumber\\
&& \hspace{1cm} +\,\,   \zeta_{a+1} \upsilon_{a+2}^\opsm \partial_{\upsilon_{a+1}^\opsm} \partial_{L_a} \partial_{L_{a+1}} +  \zeta_{a+2} \upsilon_{a+1}^\opsm   \partial_{\upsilon_{a+2}^\opsm} \partial_{L_{a+2}} \partial_{L_a}
\nonumber\\
&& \hspace{1cm} +\,\,  \half  \Big( \big(\upsilon_{a+1}^\opsm - \upsilon_a^\opsm \big)\partial_{\upsilon_{a+1}^\opsm}  + \big( \upsilon_a^\opsm - \upsilon_{a+2}^\opsm \big)\partial_{\upsilon_{a+2}^\opsm} \Big)  \zeta_{a+1} \zeta_{a+2} \partial_{L_1} \partial_{L_2} \partial_{L_3}\,.\qquad
\\
\label{man-22082022-32} && K_a^{\omsm\smfour}   =  -2\partial_{\zeta_a} + W_a\partial_{L_a} +  \alpha_{a+1 a+2}  \big( \partial_{\alpha_{aa+1}} \partial_{L_{a+2}} -  \partial_{\alpha_{a+2a}} \partial_{L_{a+1}} \big)
\nonumber\\
&&\hspace{1cm} +\,\, \half  \Big( \zeta_{a+1}(\frac{d-4}{2} - N_{\upsilonbf^\opsm})\partial_{L_{a+1}} + \zeta_{a+2} (\frac{d-4}{2} - N_{\upsilonbf^\opsm}) \partial_{L_{a+2}} \Big) \partial_{L_a} \qquad
\nonumber\\
&&\hspace{1cm} -\,\,  2\big( \zeta_{a+1} \partial_{\alpha_{aa+1}} +  \zeta_{a+2} \partial_{\alpha_{a+2a}}\big)\partial_{\upsilonbf^\opsm}
\nonumber\\
&&\hspace{1cm}  +\,\, \zeta_{a+1}\zeta_{a+2} \big( \partial_{\alpha_{a+2a}} \partial_{L_{a+1}} -  \partial_{\alpha_{aa+1}} \partial_{L_{a+2}} \big) \partial_{\upsilonbf^\opsm}\,.\qquad
\eeq
Using operators $K_a^{\omsm\smfour}$ \rf{man-22082022-32}, we now consider equations \rf{man-22082022-28}. First, we use operator $U_\zeta$ \rf{man-21082022-15} to introduce the vertex $V^\smfive$ and the operators $K_a^{\omsm\smfive}$,
\beq
\label{man-22082022-33} && V^\smfour \equiv U_\zeta V^\smfive\,, \qquad K_a^{\omsm\smfive} \equiv K_a^{\omsm\smfour}\big|_{U_\zeta}\,, \qquad a=1,2,3\,.
\\
\label{man-22082022-34} && K_a^{\omsm\smfive}  =  -2\partial_{\zeta_a}  +  \alpha_{a+1 a+2}  \big( \partial_{\alpha_{aa+1}} \partial_{L_{a+2}} -  \partial_{\alpha_{a+2a}} \partial_{L_{a+1}} \big)
\nonumber\\
&&\hspace{1cm} +\,\, \half  \Big( \zeta_{a+1} \big( \frac{d-4}{2} - N_{\upsilonbf^\opsm} \big) \partial_{L_{a+1}} + \zeta_{a+2} \big( \frac{d-4}{2} - N_{\upsilonbf^\opsm} \big) \partial_{L_{a+2}} \Big) \partial_{L_a} \qquad
\nonumber\\
&&\hspace{1cm} -\,\,  2\big( \zeta_{a+1} \partial_{\alpha_{aa+1}} +  \zeta_{a+2} \partial_{\alpha_{a+2a}}\big)\partial_{\upsilonbf^\opsm}
\nonumber\\
&&\hspace{1cm}  +\,\,  \zeta_{a+1}\zeta_{a+2} \big( \partial_{\alpha_{a+2a}} \partial_{L_{a+1}} -  \partial_{\alpha_{aa+1}} \partial_{L_{a+2}} \big) \partial_{\upsilonbf^\opsm}\,.\qquad
\eeq
In terms of the vertex $V^\smfive$, equations \rf{man-22082022-33} take the form
\be \label{man-22082022-35}
K_a^{\omsm\smfive} V^\smfive = 0 \,,  \qquad a=1,2,3\,.
\ee
Second, we use the operator $U_\zeta^\ext$ given by
\be \label{man-22082022-36}
U_\zeta^\ext = e^{u_\zeta^\ext}\,, \hspace{1cm} u_\zeta^\ext  =    \sum_{a=1,2,3} \half \zeta_a \alpha_{a+1 a+2}  \big( \partial_{\alpha_{aa+1}} \partial_{L_{a+2}} -  \partial_{\alpha_{a+2a}} \partial_{L_{a+1}} \big)\,,
\ee
and introduce the vertex $V^\smsix$ and the operators $K_a^{\omsm\smsix}$,
\beq
\label{man-22082022-37} && V^\smsix = U_\zeta^\ext V^\smfive\,, \qquad K_a^{\omsm\smsix} \equiv K_a^{\omsm\smfive}\big|_{U_\zeta^\ext}\,, \qquad a=1,2,3\,.
\\
\label{man-22082022-38} && K_a^{\omsm\smsix}  =  -2\partial_{\zeta_a} +  \half  \Big( \zeta_{a+1} \big( \frac{d-4}{2} - N_{\upsilonbf^\opsm} \big)\partial_{L_{a+1}} + \zeta_{a+2} \big( \frac{d-4}{2} - N_{\upsilonbf^\opsm} \big) \partial_{L_{a+2}} \Big) \partial_{L_a} \qquad\qquad
\nonumber\\
&&\hspace{1cm} -\,\,  2\big( \zeta_{a+1} \partial_{\alpha_{aa+1}} +  \zeta_{a+2} \partial_{\alpha_{a+2a}}\big)\partial_{\upsilonbf^\opsm}\,.
\eeq
In terms of the vertex $V^\smsix$, equations \rf{man-22082022-35} take the form
\be \label{man-22082022-39}
K_a^{\omsm\smsix} V^\smsix = 0 \,,  \qquad a=1,2,3\,.
\ee
Third, we use the operator $U_{\zeta\zeta}$ \rf{man-21082022-16} with $u_{\zeta\zeta,a}$ \rf{man-21082022-16a} to introduce  the vertex $V^\smseven$ and the operators $K_a^{\omsm\smseven}$,
\be \label{man-22082022-40}
V^\smsix = U_{\zeta\zeta} V^\smseven\,, \qquad  K_a^{\omsm\smseven}  \equiv K_a^{\omsm\smsix}\big|_{ U_{\zeta\zeta} }\,, \qquad  K_a^{\omsm\smseven}  =  -2\partial_{\zeta_a}\,.
\ee
In terms of the vertex $V^\smseven$, equations \rf{man-22082022-39} take the form
\be \label{man-22082022-41}
K_a^{\omsm\smseven} V^\smseven = 0 \,,  \qquad a=1,2,3\,.
\ee
Using the operators $K_a^{\omsm\smseven}$ \rf{man-22082022-40}, we see that equations \rf{man-22082022-41} imply that the vertex $V^\smseven$ is independent of the oscillators $\zeta_1$, $\zeta_2$, $\zeta_3$. Using the results above described, we express the vertex $p_\smp3^-$ in terms of the vertex $V^\smseven$,
\beq
\label{man-22082022-42} && p_\smp3^- = U^\ext V^\smseven\,, \qquad V^\smseven = V^\smseven(L_a, \alpha_{aa+1}, \upsilonbf^\opsm)\,,
\\
\label{man-22082022-43} &&  \hspace{1cm} U^\ext \equiv U_\beta U_{\upsilon^\omsm} U_{\upsilon^\opsm} U_{N_{\upsilon^\opsm}} U_\zeta U_\zeta^\ext U_{\zeta\zeta}\,. %
\eeq

\noinbf{ Step 5}. We consider equations \rf{man-16082022-37}.
In terms of the vertex $V^\smseven$, equations \rf{man-16082022-37} take the form
\be \label{man-22082022-44}
G_a^{\Jbf\smseven} V^\smseven = 0 \,, \hspace{1cm} G_a^{\Jbf\smseven} \equiv G_a^\Jbf\big|_{U^\ext}\,, \qquad a=1,2,3\,.
\ee
We find the following expression for the operators $G_a^{\Jbf\smseven}$:
\beq
\label{man-22082022-45} && G_a^{\Jbf\smseven}    =  G_a + \half \zeta_{a+1} L_{a+2}G_{\zeta\zeta,a+2} - \half \zeta_{a+2} L_{a+1} G_{\zeta\zeta,a+1} + \zeta_{a+1}\zeta_{a+2} G_{\zeta\zeta,a}\,,
\\
&& \hspace{1.3cm}  G_{\zeta\zeta,a}\equiv \partial_{\alpha_{a+2a}}\partial_{L_{a+1}} - \partial_{\alpha_{aa+1}}\partial_{L_{a+2}} \,,
\eeq
where the operators $G_a$ are defined in \rf{man-21082022-03}. Equations for the vertex $V^\smseven$ \rf{man-22082022-44} amount to the equations
\be \label{man-22082022-46}
G_aV^\smseven = 0\,, \qquad  G_{\zeta\zeta,a} V^\smseven =0\,, \qquad a=1,2,3\,.
\ee
In view of the relation
\be \label{man-22082022-47}
\partial_{L_{a+1}}\partial_{L_{a+2}} G_a V^\smseven = G_{\zeta\zeta,a} V^\smseven\,,
\ee
we see that the 2nd equations in \rf{man-22082022-46} follow from the 1st equations in \rf{man-22082022-46}. Using the notation
$V\equiv V^\smseven$, we cast the 1st equation in \rf{man-22082022-46} into the form given in \rf{man-21082022-03}. Using the 2nd equation in \rf{man-22082022-46} and the definition of operator $U_\zeta^\ext$  \rf{man-22082022-36}, we get the relation $U_\zeta^\ext V= V$. In view of the relation $U_\zeta^\ext U_{\zeta\zeta} V = U_{\zeta\zeta} U_\zeta^\ext V$, we then get the relation $U^\ext V = UV$, where the operator $U$ is given in \rf{man-21082022-10}. Using the relation $U^\ext V^\smseven = UV$ in \rf{man-22082022-42}, we get the vertex $p_\smp3^-$ given in \rf{man-21082022-01}.

\noinbf{Step 6}. We derive equations \rf{man-21082022-04} and \rf{man-21082022-05} and prove the equivalence of the two representations for the $u_{\zeta\zeta,a}$ given in \rf{man-21082022-16} and \rf{man-21082022-16a}.

To derive equations \rf{man-21082022-05} we note that the operator acting on the vertex $p_\smp3^-$ in \rf{man-16082022-38} is commuting with operator $U$ \rf{man-21082022-10}. Taking this into account and using the relations $N_{\zeta_a} V=0$, we see that equations \rf{man-16082022-38} lead to equations \rf{man-21082022-05}.

To derive equation \rf{man-21082022-04} we use equations \rf{man-16082022-39} and represent equation \rf{man-16082022-32} as
\be  \label{man-04092022-01}
(H - \Nbf_\zeta) p_\smp3^- = 0 \,, \qquad  H \equiv \Nbf_L    + 2N_{\upsilonbf^\opsm}  - \kbf_s + \frac{d-6}{2}\,,\qquad \Nbf_\zeta\equiv \sum_{a=1,2,3} N_{\zeta_a}\,.
\ee
Taking into account that operator $U$ \rf{man-21082022-10} is commuting with the operator $H-\Nbf_\zeta$ and using the relation $\Nbf_\zeta V=0$, we see that the equation for the vertex $p_\smp3^-$ in \rf{man-04092022-01} leads to the equation for the vertex $p_\smp3^-$ in \rf{man-21082022-04}.

To prove the equivalence of the two representations for $u_{\zeta\zeta,a}$ given in \rf{man-21082022-16} and \rf{man-21082022-16a} we use equations \rf{man-21082022-05} to represent equation \rf{man-21082022-04} as
\be \label{09072022-04}
\bigl(2\Nbf_{\upsilon^\opsm} - 2N_{\alpha_{aa+1}} - 2N_{\alpha_{a+1a+2}} - 2N_{\alpha_{a+2a}}  - \bar\kbf_s + \frac{d-6}{2} \bigr) V = 0 \,, \qquad \bar\kbf_s \equiv \sum_{a=1,2,3} \kb_{s_a}\,,
\ee
where $\kb_{s_a}$ is given in \rf{man-21082022-15}. By acting with the operator $\partial_{L_{a+1}} \partial_{L_{a+2}}$  on equation \rf{09072022-04}, we get
\be \label{09072022-06}
\big (2N_{\upsilonbf^\opsm} - \kb_{s_a}- \frac{d-6}{2} \big) \partial_{L_{a+1}}  \partial_{L_{a+2}} V = 0 \,.
\ee
Using \rf{09072022-06}, we see that the representations for $u_{\zeta\zeta,a}$ in \rf{man-21082022-16} and \rf{man-21082022-16a} are indeed equivalent.

\noinbf{Derivation of condition $\Delta_3=0$ for the vertex in \rf{man-21082022-41}, \rf{man-21082022-42}}.  Using equations \rf{man-21082022-03}-\rf{man-21082022-05}, we find the general solution for the undressed vertex,
\be \label{man-22082022-01app}
V= \upsilonbf^{\opsm\, \half(k_3 + \frac{d-6}{2})} L_1L_2   + \frac{1}{4}\Delta_3 \upsilonbf^{\opsm\, \half(k_3 + \frac{d-2}{2})} \alpha_{12}\,,\qquad  \Delta_3 = \frac{d-2}{2} - k_3\,,
\ee
where $k_3\in \No_0$. Straightforward computation gives then
\beq
\label{man-22082022-02app} && V^\omsm = \upsilon^{\half (k_3 + \frac{d-6}{2}) } \upsilon_1^{\omsm\, \half(\frac{d-2}{2}-k_3)} \upsilon_2^{\omsm\, \half(\frac{d-2}{2}-k_3)} \upsilon_3^{\omsm\, \half(k_3-\frac{d-6}{2})}  L_1 L_2
\nonumber\\
&& \hspace{0.7cm} + \,\, \frac{1}{4} \Delta_3 \upsilon^{\half (k_3 + \frac{d-2}{2})} \upsilon_1^{\omsm\, \half(\frac{d-6}{2}-k_3)} \upsilon_2^{\omsm\, \half(\frac{d-6}{2}-k_3)} \upsilon_3^{\omsm\, \half(k_3-\frac{d-2}{2})} \alpha_{12}\,.
\eeq
If $k_3\ne \frac{d-2}{2}$, then considering the $\alpha_{12}$-term in  \rf{man-22082022-02app}, we note that restriction \rf{man-21082022-09a} leads to the restrictions
\be \label{man-22082022-03app}
 d-2 \leq 2k_3 \leq d-6\,.
\ee
Restrictions \rf{man-22082022-03app} are inconsistent. We conclude therefore that $k_3= \frac{d-2}{2}$. This implies $\Delta_3=0$. Plugging $k_3=\frac{d-2}{2}$ into \rf{man-22082022-01app} and \rf{man-22082022-02app}, we see then that the vertex $V$ \rf{man-22082022-01app} satisfies the restriction \rf{man-21082022-09}, while the vertex $V^\omsm$ \rf{man-22082022-02app} satisfies the restriction \rf{man-21082022-09a}.

\appendix{ \large Derivation of densities
\rf{man-16082022-40}, \rf{man-16082022-41} }

We prove the following {\bf Statement}: If cubic vertex $p_\smp3^-$ satisfies equations \rf{man-16082022-31}, \rf{man-16082022-33}-\rf{man-16082022-36}, then equations \rf{man-17122016-19}, \rf{man-17122016-20} lead to the densities $j_\smp3^{-i}$ and $k_\smp3^-$ given in \rf{man-16082022-40}, \rf{man-16082022-41}, while equation \rf{man-17122016-21} does not impose additional restrictions.

\noinbf{Proof of \rf{man-16082022-40}}. Using equations \rf{man-16082022-35}, \rf{man-16082022-36} and relations \rf{man-16082022-06} , \rf{man-16082022-22}, \rf{man-16082022-25}, we note the following relation:
\be \label{man-06092022-01}
\Kbf^i p_\smp3^-  =  - \sum_{a=1,2,3}\frac{2\betach_a}{3\beta_a} \alpha_a^i\partial_{L_a} p_\smp3^- \,.
\ee
Using equation \rf{man-17122016-19} and relation \rf{man-06092022-01}, we get the density $j_\smp3^{-i}$ given in  \rf{man-16082022-40}.

\noinbf{Proof of \rf{man-16082022-41}}. Plugging the density $j_\smp3^{-i}$ \rf{man-16082022-40} into \rf{man-17122016-20}, we represent equation  \rf{man-17122016-20} as
\be \label{man-06092022-02}
\Kbf^i \sum_{a=1,2,3}\frac{2\betach_a}{3\beta_a} \alpha_a^j\partial_{L_a}
p_\smp3^-  + \frac{\Delta_\beta}{9} \partial_{\Po^i}\partial_{\Po^j} p_\smp3^- +  \delta^{ij} \Big( \frac{1}{6}  \Mbf^\ommsm - \frac{\Delta_\beta}{18}  \partial_{\Po^l}\partial_{\Po^l} \Big) p_\smp3^-   = \delta^{ij} k_\smp3^-\,.
\ee
We consider the $\Kbf^i$-term in \rf{man-06092022-02}. Using the expression for $\Kbf^i$ \rf{man-17122016-22} and relation \rf{man-06092022-01}, we find the relation
\beq
\label{man-06092022-03} && \Kbf^i\frac{\betach_a}{\beta_a} \alpha_a^j\partial_{L_a}   p_\smp3^- =\delta^{ij} \big( \frac{\betach_a}{\beta_a} m_a^\omsm + \frac{\betach_a^2}{3\beta_a} \alpha_a^l\partial_{\Po^l}  \big)\partial_{L_a} p_\smp3^- + K_a^{ij} p_\smp3^-\,,
\\
\label{man-06092022-04} && K_a^{ij} \equiv - \frac{\betach_a^2}{3\beta_a} \alpha_a^i \sum_{b=1,2,3} \frac{1}{\beta_b} \alpha_b^j\partial_{L_b}\partial_{L_a}
\nonumber\\
&& \hspace{0.7cm} +\,\, \frac{2\beta}{\beta_a^2} \alpha_a^j \sum_{b=1,2,3} \frac{1}{\beta_b}\alpha_b^i\partial_{L_b}\partial_{L_a} + \frac{\betach_a}{\beta_a} \alpha_a^j \sum_{b=1,2,3} \frac{\betach_a-\betach_b}{3\beta_b}\alpha_b^i \partial_{L_a} \partial_{L_b}\,.
\eeq
Before to proceed we mention two helpful relations.
\be \label{man-06092022-05}
\partial_{L_a}^2 p_\smp3^- = 0 \,, \qquad \partial_{\Po^i} p_\smp3^- =\sum_{a=1,2,3}\frac{1}{\beta_a}\alpha_a^i\partial_{L_a} p_\smp3^-\,.
\ee
The 1st relation in \rf{man-06092022-05} is valid for the cubic vertex of scalar and vectors fields in view of equations \rf{man-16082022-38}, while the 2nd relation in \rf{man-06092022-05} is obtained by using the definition of $L_a$ in \rf{man-16082022-03}. Using relations \rf{man-06092022-05} and the notation
\be  \label{man-06092022-06}
\Kbf^{ij} \equiv \sum_{a=1,2,3} K_a^{ij}\,,
\ee
we find the following helpful relations:
\beq
\label{man-06092022-07} && \Kbf^{ij} p_\smp3^-  = -  \sum_{a=1,2,3} \frac{\Delta_\beta}{6\beta_{a+1}\beta_{a+2}} \big( \alpha_{a+1}^i\alpha_{a+2}^j + \alpha_{a+1}^j\alpha_{a+2}^i \big) \partial_{L_{a+1}} \partial_{L_{a+2}} p_\smp3^- \,,
\\
\label{man-06092022-08}  && \partial_{\Po^i} \partial_{\Po^j}  p_\smp3^-  = \sum_{a=1,2,3} \frac{1}{\beta_{a+1}\beta_{a+2}}(\alpha_{a+1}^i \alpha_{a+2}^j + \alpha_{a+1}^j \alpha_{a+2}^i \big) \partial_{L_{a+1}} \partial_{L_{a+2}}  p_\smp3^- \,,
\\
\label{man-06092022-09} && \partial_{\Po^l} \partial_{\Po^l}  p_\smp3^-  = \sum_{a=1,2,3} \frac{2}{\beta_{a+1}\beta_{a+2}} \alpha_{a+1a+2} \partial_{L_{a+1}} \partial_{L_{a+2}}  p_\smp3^- \,.
\eeq
Using relations \rf{man-06092022-03} and \rf{man-06092022-06}, we find for the $\Kbf^i$-term in \rf{man-06092022-02},
\be \label{man-06092022-10}
\Kbf^i \sum_{a=1,2,3} \frac{2\betach_a}{3\beta_a} \alpha_a^j\partial_{L_a}
p_\smp3^-    = \delta^{ij} \sum_{a=1,2,3}\big( \frac{2\betach_a}{3\beta_a} m_a^\omsm + \frac{2\betach_a^2}{9\beta_a} \alpha_a^l\partial_{\Po^l}  \big)\partial_{L_a} p_\smp3^- + \frac{2}{3} \Kbf^{ij}p_\smp3^-\,.
\ee
Using relations \rf{man-06092022-07}, \rf{man-06092022-08}, and \rf{man-06092022-10}, we note the cancellation of the $\alpha^i\alpha^j$-terms in \rf{man-06092022-02} and get the relation
\be
\label{man-06092022-11}  k_\smp3^- =   \sum_{a=1,2,3} \Big( \frac{2\betach_a}{3\beta_a} m_a^\omsm \partial_{L_a} + \frac{2\betach_a^2}{9\beta_a} \alpha_a^l\partial_{\Po^l} \partial_{L_a} \Big) p_\smp3^-  - \frac{\Delta_\beta}{18}\partial_{\Po^i}\partial_{\Po^i} p_\smp3^- +    \frac{1}{6} \Mbf^\ommsm p_\smp3^-\,.
\ee
Using relations \rf{man-06092022-05},  \rf{man-06092022-09}, and \rf{man-06092022-11}, we get the relation
\beq
\label{man-06092022-12} && k_\smp3^- =   \sum_{a=1,2,3} \Big( \frac{2\betach_a}{3\beta_a} m_a^\omsm \partial_{L_a} + \frac{2}{9\beta_{a+1}\beta_{a+2}} \big( \betach_{a+1}^2 + \betach_{a+2}^2\big) \alpha_{a+1a+2} \partial_{L_{a+1}} \partial_{L_{a+2}} \Big) p_\smp3^-
\nonumber\\
&& \hspace{0.7cm} + \,\,   \frac{1}{6} \Mbf^\ommsm p_\smp3^- -  \sum_{a=1,2,3} \frac{\Delta_\beta}{9\beta_{a+1}\beta_{a+2}} \alpha_{a+1a+2} \partial_{L_{a+1}} \partial_{L_{a+2}} p_\smp3^-\,.
\eeq
Finally, using the relation
\be  \label{man-06092022-14}
2\betach_{a+1}^2 + 2\betach_{a+2}^2 - \Delta_\beta = 8\beta_a^2 - 2\beta_{a+1} \beta_{a+2}
\ee
in \rf{man-06092022-12}, we get the density $k_\smp3^-$ given in \rf{man-16082022-41}.

\noinbf{Check of equation \rf{man-17122016-21}}. Using $j_\smp3^{-i}$, $k_\smp3^-$ given in \rf{man-16082022-40},  \rf{man-16082022-41}, we find the relations
\beq
\label{04082022-02} &&  \Kbf^i k_\smp3^-    +  \Kbf_\Xbf^{il} j_\smp3^{-l}     = C_1^i + C_2^i + C_3^i + C_4^i\,,
\\
\label{04082022-02a1} && \hspace{1cm} C_1^i \equiv  \Kbf^i \sum_{a=1,2,3}  \frac{2\betach_a}{3\beta_a}m_a^\omsm \partial_{L_a} p_\smp3^- - \frac{1}{3} \Mbf^\ommsm  \sum_{a=1,2,3}  \frac{2\betach_a}{3\beta_a} \alpha_a^i \partial_{L_a} p_\smp3^-\,,
\\
\label{04082022-02a2} && \hspace{1cm} C_2^i \equiv \Kbf^i \sum_{a=1,2,3} f_a\alpha_{a+1a+2} \partial_{L_{a+1}}  \partial_{L_{a+2}}  p_\smp3^-\,, \hspace{1cm}  f_a \equiv \frac{2}{9} \big( \frac{4\beta_a^3}{\beta}-1 \big)\,,\qquad
\\
\label{04082022-02a3} && \hspace{1cm} C_3^i \equiv   \frac{1}{6} [\Kbf^i,\Mbf^\ommsm] p_\smp3^-\,,
\\
\label{04082022-02a4} && \hspace{1cm} C_4^i \equiv - \frac{\Delta_\beta}{9}\big( \partial_{\Po^i} \partial_{\Po^j}
- \half \delta^{ij} \partial_{\Po^l} \partial_{\Po^l} \big) \sum_{a=1,2,3}  \frac{2\betach_a}{3\beta_a} \alpha_a^j \partial_{L_a} p_\smp3^-\,.
\eeq
We now present the result of our computation of $C_1^i$, $C_2^i$, $C_3^i$, and $C_4^i$,
\beq
\label{04082022-03} && C_1^i  = \big( X_\Ism^i + X_\IIsm^i + Y_\Ism^i \big)p_\smp3^-\,, \hspace{1.2cm}  C_2^i =     \big( X_\IIIsm^i + Y_\IIsm^i + Y_\IIIsm^i \big)p_\smp3^-\,,
\\
\label{04082022-05} && C_3^i  = \big( X_\IVsm^i + Y_\IVsm^i\big)p_\smp3^-\,, \hspace{2cm} C_4^i = Y_\Vsm^i p_\smp3^-\,,
\\
\label{04082022-08} && X_\Ism^i \equiv  \sum_{a=1,2,3} \frac{2}{3} \big(\frac{2\beta}{\beta_{a+1}\beta_{a+2}^2} - \frac{\betach_{a+1}\betach_{a+2}}{3\beta_{a+1}\beta_{a+2}}\big) \alpha_{a+1}^i  m_{a+2}^\omsm \partial_{L_{a+1}} \partial_{L_{a+2}}\,,
\\
\label{04082022-09} && \hspace{0.6cm}+\,\, \sum_{a=1,2,3} \frac{2}{3} \big(\frac{2\beta}{\beta_{a+1}^2\beta_{a+2}} - \frac{\betach_{a+1}\betach_{a+2}}{3\beta_{a+1}\beta_{a+2}}\big) \alpha_{a+2}^i  m_{a+1}^\omsm \partial_{L_{a+1}} \partial_{L_{a+2}}\,, \qquad
\\
\label{04082022-10} && X_\IIsm^i \equiv  - \sum_{a=1,2,3} \frac{4}{9\beta_{a+1}\beta_{a+2}} \Big( \betach_{a+1}^2 \alpha_{a+1}^i m_{a+2}^\omsm  + \betach_{a+2}^2 \alpha_{a+2}^i m_{a+1}^\omsm \Big) \partial_{L_{a+1}} \partial_{L_{a+2}}\,,\qquad
\\
\label{04082022-11} && X_\IIIsm^i\equiv \sum_{a=1,2,3} f_a \big(\alpha_{a+1}^i m_{a+2}^\omsm +  \alpha_{a+2}^i m_{a+1}^\omsm \big) \partial_{L_{a+1}} \partial_{L_{a+2}}\,,
\\
\label{04082022-12} && X_\IVsm^i \equiv \sum_{a=1,2,3} \frac{2\Delta_\beta}{9\beta_{a+1}\beta_{a+2}} \big(   \alpha_{a+1}^i m_{a+2}^\omsm +  \alpha_{a+2}^i m_{a+1}^\omsm \big)
\partial_{L_{a+1}} \partial_{L_{a+2}}\,,
\\
\label{04082022-13} && Y_\Ism^i \equiv - \sum_{a=1,2,3} \frac{8\betach_a}{9\beta_a} \alpha_a^i \alpha_{a+1a+2} \partial_{L_1} \partial_{L_2} \partial_{L_3}
\\
\label{04082022-14} &&   Y_\IIsm^i \equiv\sum_{a=1,2,3}  \frac{2\betach_a}{27\beta_a}\big(\frac{8\beta_a^3}{\beta} +1\big) \alpha_a^i \alpha_{a+1a+2} \partial_{L_1} \partial_{L_2}\partial_{L_3}\,,
\\
\label{04082022-15} &&  Y_\IIIsm^i \equiv  - \sum_{a=1,2,3} \frac{4\betach_a\Delta_\beta}{9\beta} \alpha_a^i\alpha_{a+1a+2} \partial_{L_1} \partial_{L_2}\partial_{L_3}\,,
\\
\label{04082022-16} && Y_\IVsm^i = -\frac{1}{4} Y_\Ism^i\,, \qquad Y_\Vsm^i = -\frac{1}{3} Y_\IIIsm^i\,.
\eeq
Using the relations
\be
X_\Ism^i + X_\IIsm^i + X_\IIIsm^i + X_\IVsm^i = 0 \,, \qquad Y_\Ism^i + Y_\IIsm^i + Y_\IIIsm^i + Y_\IVsm^i + Y_\Vsm^i=0 \,,
\ee
we see that equation \rf{man-17122016-21} is satisfied.

\noinbf{Comments on derivation of $C_1^i$, $C_3^i$ \rf{04082022-03}, \rf{04082022-05}}. The $C_1^i$ and $C_3^i$ are defined in \rf{04082022-02a1}, \rf{04082022-02a3}. Derivation of the expression for $C_1^i$ and $C_3^i$ given in \rf{04082022-03}, \rf{04082022-05} requires the knowledge of some additional relations which we now discuss. First, we consider $C_1^i$. We note the relations
\beq
\label{04082022-26} && \Kbf^i \sum_{a=1,2,3} \frac{2\betach_a}{3\beta_a} m_a^\omsm \partial_{L_a} p_\smp3^- = X_\Ism^i +  \sum_{a=1,2,3} \frac{2\betach_a}{3\beta_a} \alpha_a^i M_a^\ommsm \partial_{L_a}p_\smp3^-\,,
\\
\label{04082022-27} && C_1^i = X_\Ism^i p_\smp3^-  + C_{1,\add}^i\,, \qquad  C_{1,\add}^i \equiv \sum_{a=1,2,3}  \frac{2\betach_a}{3\beta_a}  \alpha_a^i \partial_{L_a} \big( M_a^\ommsm   - \frac{1}{3} \Mbf^\ommsm \big) p_\smp3^-\,,
\eeq
where $C_1^i$ \rf{04082022-27} is obtained by using \rf{04082022-02a1} and \rf{04082022-26}. Now we consider $C_{1,\add}^i$, \rf{04082022-27}. We make the following important observation.
The relation for the vertices $p_\smp3^-$ and $V^\smone$ given in \rf{man-22082022-01} implies that, in terms of $p_\smp3^-$, equations \rf{man-22082022-14} can be represented as
\beq
\label{04082022-28}  && K_1 p_\smp3^-  = K_2 p_\smp3^-  = K_3 p_\smp3^- \,,\qquad K_a\equiv K_a^{\ommsm\smone}\big|_{U_\beta^{-1}}\,, \qquad  K_a = M_a^\ommsm + X_a\,,
\\
\label{04082022-30} && X_a \equiv  \frac{\betach_a}{\beta_a} m_a^\omsm \partial_{L_a} + m_{a+1}^\omsm \partial_{L_{a+1}}  - m_{a+2}^\omsm \partial_{L_{a+2}} + 2 \alpha_{a+1a+2}\partial_{L_{a+1}} \partial_{L_{a+2}}\,.\qquad
\eeq
It is the relations for $p_\smp3^-$ in \rf{04082022-28} that are the additional relations we need to get $C_1^i$, $C_3^i$ given in  \rf{04082022-03}, \rf{04082022-05}. Namely, the relations for $p_\smp3^-$ in \rf{04082022-28} lead to the relation
\be \label{04082022-31}
\sum_{a=1,2,3}  \frac{2\betach_a}{3\beta_a}  \alpha_a^i  \partial_{L_a} \big( K_a  - \frac{1}{3}\Kbf\big) p_\smp3^-  = 0 \,, \qquad \Kbf\equiv \sum_{a=1,2,3} K_a\,. %
\ee
Using relation \rf{04082022-31}, the definition of $C_{1,\add}^i$ in \rf{04082022-27}, and the expression for the operators $K_a$ in \rf{04082022-28}, we get the relation
\beq
\label{04082022-32}  &&  C_{1,\add}^i  =  - \sum_{a=1,2,3}  \frac{2\betach_a}{3\beta_a}  \alpha_a^i  \partial_{L_a} \big( X_a - \frac{1}{3} \Xbf\big)p_\smp3^- \,,\qquad \Xbf \equiv \sum_{a=1,2,3} X_a\,.
\eeq
Plugging expression for $X_a$ \rf{04082022-30} into \rf{04082022-32}, we get $C_{1,\add}^i = (X_\IIsm^i + Y_\Ism^i)p_\smp3^-$. Plugging such $C_{1,\add}^i$ into the 1st relation in \rf{04082022-27}, we get expression for $C_1^i$ given in \rf{04082022-03}.

Second, we consider $C_3^i$. Using operator $\Kbf^i$ \rf{man-16082022-06}, we get the relation
\be \label{04082022-36}
\frac{1}{6} [\Kbf^i, \Mbf^\ommsm] = - \sum_{a=1,2,3} \frac{1}{3\beta_a}\alpha_a^i\partial_{L_a} \Mo^\ommsm \,, \qquad \Mo^\ommsm\equiv \frac{1}{3} \sum_{a=1,2,3} \betach_a M^\ommsm_a\,.
\ee
From relations \rf{04082022-02a3}, \rf{04082022-36}, we see that in order to find $C_3^i$ we should find action of the operator $\Mo^\ommsm$ on the vertex $p_\smp3^-$. To this end, using the relations for $p_\smp3^-$ in \rf{04082022-28}, we note the relations
\beq
\label{04082022-36man1}  && \Ko p_\smp3^-  = 0 \,, \qquad \Ko\equiv \frac{1}{3} \sum_{a=1,2,3} \betach_a K_a\,,
\\
\label{04082022-36man2} && \Mo^\ommsm p_\smp3^-  = -\Xo p_\smp3^- \,, \qquad \Xo\equiv \frac{1}{3} \sum_{a=1,2,3} \betach_a X_a\,,
\eeq
where the relation for the vertex $p_\smp3^-$ in \rf{04082022-36man2} is obtained from the relation for the vertex $p_\smp3^-$ in \rf{04082022-36man1} and the expression for $K_a$ in \rf{04082022-28}. Plugging expression for $X_a$ \rf{04082022-30} into \rf{04082022-36man2}, we get the relation
\be \label{04082022-37}
\Mo^\ommsm p_\smp3^-  =  - \sum_{a=1,2,3} \big( \frac{2\Delta_\beta}{3\beta_a} m_a^\omsm \partial_{L_a} + \frac{2}{3}\betach_a \alpha_{a+1a+2}\partial_{L_{a+1}} \partial_{L_{a+2}} \big) p_\smp3^-\,.
\ee
Using relations \rf{04082022-36}, \rf{04082022-37}, and \rf{04082022-02a3}, we get $C_3^i$ \rf{04082022-05}.

\end{document}